\documentclass[a4paper,twocolumn,prb,aps,10pt]{revtex4-1}

\usepackage{amsmath}
\usepackage{graphicx}
\usepackage{textcomp}
\usepackage{hyperref}							
\hypersetup{colorlinks=true,allcolors=blue}

\newcommand{\nn}{\nonumber\\}
\newcommand{\ddt}{\frac{\partial}{\partial t}}
\newcommand{\ddtau}{\frac{\partial}{\partial\tau}}
\renewcommand{\Im}{\textrm{Im}}

\begin{document}
\title{
Comparison of different concurrences 
characterizing  photon-pairs generated
in the biexciton cascade in quantum dots
coupled to microcavities
}
\author{M. Cygorek}
\affiliation{Department of Physics, University of Ottawa, ON K1N 6N5 Ottawa, Canada}
\author{F. Ungar}
\affiliation{Lehrstuhl f{\"u}r Theoretische Physik III, Universit{\"a}t Bayreuth, 95440 Bayreuth, Germany}
\author{T. Seidelmann}
\affiliation{Lehrstuhl f{\"u}r Theoretische Physik III, Universit{\"a}t Bayreuth, 95440 Bayreuth, Germany}
\author{A. M. Barth}
\affiliation{Lehrstuhl f{\"u}r Theoretische Physik III, Universit{\"a}t Bayreuth, 95440 Bayreuth, Germany}
\author{A. Vagov}
\affiliation{Lehrstuhl f{\"u}r Theoretische Physik III, Universit{\"a}t Bayreuth, 95440 Bayreuth, Germany}
\author{T. Kuhn}
\affiliation{Institut f{\"u}r Festk{\"o}rpertheorie, Universit{\"a}t M{\"u}nster, 48149 M{\"u}nster, Germany}
\author{V. M. Axt}
\affiliation{Lehrstuhl f{\"u}r Theoretische Physik III, Universit{\"a}t Bayreuth, 95440 Bayreuth, Germany}
\begin{abstract}

We compare three different notions of concurrence to measure the polarization entanglement
of two-photon states generated by the biexciton cascade in a quantum dot embedded in a microcavity. 
The focus of the paper lies on the often-discussed situation of a dot with finite biexciton binding energy in a cavity 
tuned to the two-photon resonance.
Apart from the time-dependent concurrence, which can be assigned to the two-photon density matrix at any point in time,
we study single- and double-time integrated concurrences commonly used in the literature that are based on
different quantum state reconstruction schemes.
In terms of the photons detected in coincidence measurements, we argue that the single-time integrated concurrence
can be thought of as the concurrence of photons simultaneously emitted from the cavity without resolving the common 
emission time, while the more widely studied double-time integrated concurrence refers to photons that are neither filtered 
with respect to the emission time of the first photon nor with respect to the delay time between the two emitted photons.
Analytic and numerical calculations reveal that  the single-time integrated concurrence indeed agrees well with the 
typical value of the time-dependent concurrence at long times, even when the interaction between the quantum dot and 
longitudinal acoustic phonons is accounted for. 
Thus, the more easily measurable single-time integrated concurrence gives access to the physical information represented 
by the time-dependent concurrence.
However, the double-time integrated concurrence shows a different behavior with respect to changes in the exciton 
fine structure splitting and even displays a completely different trend when the ratio between the cavity loss rate and 
the fine structure splitting is varied while keeping their product constant.
This implies the non-equivalence of the physical information contained in the time-dependent and single-time integrated 
concurrence on the one hand and the double-time integrated concurrence on the other hand.

\end{abstract}
\maketitle
\section{Introduction}

Many applications in quantum communication require the generation of
entangled photon pairs \cite{Quantum_cryptography_RevModPhys,
Photonic_quantum_technologies, Orieux_entangled,Zeilinger_entangled,
ncomm_dalacu2014, entangled-photon1}. 
One particularly promising 
way of producing such pairs consists of using the biexciton 
cascade in quantum dots \cite{entangled-photon2,
Stevenson2006,Hudson2007,Hafenbrak,StevensonPRL2008,Bounouar_2018,Pfanner2008,
BiexcCascade_dressed,Orieux_entangled}, 
which can be sketched roughly as follows:
An initially prepared biexciton state of a quantum dot decays to
one of two possible single exciton states while emitting a photon.
A second photon can be emitted while the quantum dot
relaxes to its ground state. Because of the optical 
selection rules, the two different paths lead to an emission of 
either two horizontally or two vertically polarized photons.
As the biexciton decay is a quantum mechanical process the system will, in
general, be in a superposition of states from both paths.
If both paths are completely symmetric, one expects that there is a high probability
to find the system
in the fully entangled state 
$|\psi^+\rangle=\frac 1{\sqrt 2}\big(|HH\rangle +|VV\rangle\big)$,
where $|HH\rangle$ ($|VV\rangle$) denotes the state where the quantum dot
is in its ground state and two horizontally (vertically) polarized photons
have been emitted. 

However, the situation in real quantum dots often deviates from the 
ideal picture described above.
First of all, the exchange interaction typically introduces 
an energetic splitting between the two excitonic states on the order of
several tens to hundreds of {\textmu}eV 
\cite{entangled-photon2,Biexc_FSS_electrical_control,Biexc_FSS_electrical_control_Bennett}. 
Thus, the two paths become asymmetric, leading to a deviation from the
usually desired state $|\psi^+\rangle$.
Besides effects related to the fine structure splitting another source for
deviations from the ideal situation are environment couplings, in particular
to phonons. These couplings lead to decoherence and relaxation and are the reason why
the system has to be represented by a mixed rather than a pure state.
These detrimental effects can be suppressed by engineering the 
quantum dot devices accordingly. For example, the fine structure splitting 
between the excitonic states can be reduced by applying electrical 
\cite{Biexc_FSS_electrical_control,Biexc_FSS_electrical_control_Bennett} 
or strain fields \cite{Biexc_FSS_strain_control_Seidl}
or by growing quantum dots within highly symmetric structures such as
nanowires \cite{nanowireQD_Huber}.
Here, we consider another approach to obtain more symmetric paths, which 
is achieved by embedding the quantum dot in a microcavity. Then,
the coupling between the electronic states in the dot and the cavity modes 
leads to an overall faster dynamics, which reduces the time available for
dephasing processes. Furthermore, tuning the cavity modes to the two-photon
resonance between the ground and the biexciton state of the dot enhances
two-photon processes that are much less affected by the splitting of the
excitonic states than successive single-photon processes 
\cite{delValle_twoPhoton,Jahnke2012}. 

The wide interest in entanglement is twofold: on the one hand 
the occurrence of entanglement is one of the
key differences between classical and quantum physics and on the other 
hand it has practical implications as it provides new ways of control
as needed, e.g., for establishing secure quantum communication protocols \cite{liao:18}. 
The essence of the control aspect is that by performing a measurement on one part of the
system one determines the outcome of measurements on another part of the system which otherwise 
would have been undetermined. If, e.g., the system is in the maximally entangled 
(non-factorizable) state $|\psi^+\rangle$ and one detects the polarization of one of the photons to be $H$,
the state collapses into the factorized state $|HH\rangle$ and a polarization
measurement on the second photon will necessarily yield $H$.

In order to compare two arbitrary states with respect to the amount of control 
obtainable by performing measurements as described above, one needs a measure of
entanglement. For a pure state $|\Psi\rangle$ in a bipartite system with density matrix $\rho$ and parts $A$ and $B$
with reduced density matrices 
$\rho_{A}= \mathrm{Tr}_{B}\rho$ and
$\rho_{B}=\mathrm{Tr}_{A}\rho$, respectively,
it is common to define the entanglement $E(|\Psi\rangle)$ using the von-Neumann entropies
of the subsystems \cite{bennett96}:
\begin{align}
  E(|\Psi\rangle) 
= -\mathrm{Tr}_{A}\Big(\rho_{A}\log_{2}\rho_{A}\Big)
= -\mathrm{Tr}_{B}\Big(\rho_{B}\log_{2}\rho_{B}\Big),
\label{E-def-pure}
\end{align}
where the second equality in Eq.~\eqref{E-def-pure} follows from the Schmidt decomposition
\cite{nielsen}. This entropy  represents the missing information about a subsystem because of 
its entanglement with the other. Performing a measurement  on one of the subsystems
that collapses the system state into a factorizable state causes the subsystem entropy to drop to zero
such that the previously missing information $E(|\Psi\rangle)$ is recovered. Therefore, $E(|\Psi\rangle)$
can also be considered to be a measure of the possible amount of control over subsystem $A$ by performing
measurements on $B$.

For a mixed state missing information on a subsystem can arise due to its entanglement with the remaining
part of the system as well as because of the ensemble averaging.
There are a number of proposals to identify the corresponding contribution resulting from entanglement 
\cite{bennett96,Wootters:2001}. Probably the most common proposal is the entanglement of formation. 
For a decomposition 
\begin{align}
  \rho = \sum_{j} p_{j}\,|\Psi_{j}\rangle\langle\Psi_{j}|
\label{decomp}
\end{align}
of a density matrix $\rho$ with probabilities $p_{j}$ and not necessarily orthogonal states $|\Psi_{j}\rangle$
one assigns the entanglement:
\begin{align}
 E(\rho,\{p_{j},|\Psi_{j}\rangle \} )  = \sum_{j} p_{j}\,E(|\Psi_{j}\rangle).
\end{align}
Then the entanglement of formation is defined as:
\begin{align}
  E_{f}(\rho) = \inf\, E(\rho,\{p_{j},|\Psi_{j}\rangle \} ),
\end{align}
where the infimum is taken over all possible decompositions in the form of Eq.~\eqref{decomp}. 
Thus, $E_{f}(\rho)$ represents the amount of pure-state entanglement that is at least present in a mixed state
described by a given density matrix $\rho$.
The entanglement of formation is particularly attractive because, unlike most other proposed measures of entanglement, 
it can be evaluated directly from the elements of the density matrix. 
To this end usually the concurrence $C$ is introduced which is related to the entanglement of
formation by
\begin{align}
  E_{f}(\rho) = \mathcal{E}\big(C(\rho)\big),
\label{Ef-C}
\end{align}
where $\mathcal{E}$ is monotonically increasing for $0\le C \le 1$ (cf.~Ref.~\onlinecite{Wootters:2001}
for an explicit expression for $\mathcal{E}$). Due to the monotonicity of $\mathcal{E}$
the concurrence is a measure of the entanglement of formation in its own right. Although 
the concurrence is less intuitive than the entanglement of formation and its
physical interpretation derives only from its relation to $E_{f}(\rho)$, it is particularly attractive
for practical applications because, as shown by Wootters \cite{Wootters1998}, it 
can be easily calculated from the elements of $\rho$ without having to perform a search for the infimum over
all possible decompositions of $\rho$. 

In the case of the biexciton cascade
where no direct transitions between the two exciton states occur, 
the (unnormalized) density matrix in the two-photon subspace
\begin{align}
\label{eq:twophotondensmat}
\rho_{ij,kl}(t)=\langle a^\dagger_i(t) a^\dagger_j(t) a_k(t) a_l(t)\rangle,
\end{align}
with $a^\dagger_i(t)$ and $a_l(t)$ being the cavity photon creation and annihilation
operators with polarization directions $i,j,k,l\in\{H,V\}$ in the Heisenberg
picture, has only four non-vanishing elements, namely
$\rho_{HH,HH}, \rho_{VV,VV}, \rho_{HH,VV}$, and $\rho_{VV,HH}$.
Since we are dealing with a system where the photon number is not conserved,
$\rho_{ij,kl}(t)$ has not unit trace for all times. The non-vanishing elements
of the normalized density matrix can be represented accordingly by:
\begin{align}
  \rho^{N}_{j,l}(t) = \frac{\rho_{jj,ll}(t)}{\rho_{HH,HH}(t)+\rho_{VV,VV}(t)}.
\end{align}
The general expression for the concurrence\cite{Wootters1998} then reduces to the normalized
coherences between the states $|HH\rangle$ and $|VV\rangle$, which correspond to
horizontal $H$ or vertical $V$ polarization of the emitted photons, i.e.:
\begin{align}
\label{eq:C}
&C(t)=2|\rho^{N}_{H,V}(t)|.
\end{align}
Thereby, the concurrence relates two conceptually distinct properties of 
quantum systems:
entanglement, which specifies how much the measurement of one qubit influences
the measurement outcome of the second qubit, and coherence, which
determines, e.g., the visibility of interference effects. 
A theoretical study where the time-dependent concurrence, which is assigned to 
the density matrix in the two-photon subspace at a given time, has been used as a figure of merit for
entanglement has been performed, e.g., in Ref.~\onlinecite{hein2014}.
However, in contrast to the investigation here, in Ref.~\onlinecite{hein2014}
the coupling of the dot to a continuum of half-space photon modes has been considered.

Although the elements of the density matrix are in principle all observable,
it is often difficult to resolve their full time dependence experimentally and thereby
determine the time-dependent concurrence given by Eq.~\eqref{eq:C}.
To reconstruct the two-photon density matrix from experimental data, 
one usually uses quantum state tomography, a technique based on polarization-dependent 
photon coincidence measurements \cite{QuantumStateTomography,Troiani2006}.
Because these coincidence measurements typically give only information about 
the polarization degree of freedom and the time delay $\tau$ between the 
two measured photons, but do not resolve the time $t$ of the first photon count
with respect to the preparation of the biexciton state ($t=0$), 
in such experiments one only has access to quantities integrated over the time $t$.
Therefore, studies of the photon pairs generated
via the biexciton cascade often define figures of merit, which are then also
called concurrence, but replace the density matrix
in Eq.~\eqref{eq:C} by the respective expressions 
obtained from the quantum state reconstruction.
The latter involve in general averages over $t$ and $\tau$.
Most often discussed is the limiting case of long 
averaging intervals for both times \cite{Jahnke2012,Heinze2017,Pfanner2008}.
However, also the case where the averaging window for $t$ is infinite, 
while for $\tau$ the limit of a vanishing averaging interval 
is approached, is experimentally accessible \cite{StevensonPRL2008, Bounouar_2018}
and has been studied theoretically \cite{carmele11,EdV}.

That these figures of merit may indeed differ substantially because of the
differences in time averaging
can be readily seen from the following argument due to Stevenson \emph{et al.}
\cite{StevensonPRL2008}.
Consider a biexciton cascade where an initially prepared biexciton 
decays to an exciton while emitting a single photon. 
A second photon is emitted after a delay time $\tau$, during which 
the different exciton states aquire a phase difference 
$\phi=\tau\delta/\hbar$ due to the fine structure splitting $\delta$. 
Disregarding environment influences which lead to a mixed photon state
and concentrating only on the free time evolution the resulting two-photon
state is $|\Psi\rangle=\frac 1{\sqrt{2}}\big( |HH\rangle +
e^{i\tau\delta/\hbar}|VV\rangle\big)$.
Obviously, $|\Psi\rangle$ is a maximally entangled state at each point in time
for any given delay.
If, however, measurements are performed that do not discriminate between different
delay times $\tau$ of the emission, the effective time-integration leads to
significant cancellations of the phases $e^{i\tau\delta/\hbar}$. 
Stevenson \emph{et al.}\cite{StevensonPRL2008} 
have performed experiments where the probability for finding the maximally entangled
state $|\Psi\rangle=\frac 1{\sqrt{2}}(|HH\rangle+|VV\rangle)$ 
in the two-photon state 
generated via the biexciton cascade has been determined
as a function of the integration window for the delay time $\tau$. 
Indeed, it was found that this probability significantly drops the longer
the $\tau$ sampling interval is taken.
Thus, filtering photon pairs with nearly equal emission times
reveals a high degree of entanglement while measurements involving long $\tau$ sampling times
indicate a much lower entanglement.
A similar experimental analysis has been recently performed by Bounouar \emph{et al.}\cite{Bounouar_2018}
where it was concluded that the main limit of entanglement fidelity is the time resolution
in the experiment.

The goal of this article is to compare the 
definitions of concurrences 
commonly used in the literature 
involving either
single- or double-time averages 
with the time-dependent concurrence given by Eq.~\eqref{eq:C}.
To be specific, we study the case of the biexciton cascade in a quantum dot
inside a microcavity.
Concentrating at first on a model without phonons, we derive analytic expressions for the different concurrences for a quantum
dot with finite biexciton binding energy in a cavity tuned to the two-photon
resonance, a configuration which was already found to be favorable for a high 
degree of polarization entanglement\cite{Jahnke2012}.
The analytic results are valid also beyond the weak-coupling limit and agree 
qualitatively with numerical calculations.
We find that the concurrence
based on a single-time integrated two-photon density matrix yields very similar
results as the time-dependent concurrence in Eq.~\eqref{eq:C}.
It is therefore possible to access the information represented by the time-dependent
concurrence, i.e., the entanglement of formation contained in the so prepared state
of the cavity photons,
by recording the more easily measurable single-time integrated concurrence.
This remains true even when the interaction between the quantum dot states and
longitudinal acoustic phonons is taken into account in numerical calculations.
However, this information cannot be accessed by measuring
two-time integrated correlation functions since it turns out
that the latter exhibit quantitatively and qualitatively different
dependencies on parameters like the fine structure splitting.
It is most striking that when comparing single- and double-time integrated concurrences 
even trends reverse, such as the dependence on the cavity loss rate in the presence
of phonons.
Furthermore, already without phonons, these two quantities show a completely reversed 
trend when the ratio between the cavity loss rate and the fine structure splitting 
is varied while their product is kept constant.
This leads to the conclusion that the single-time integrated and the double-time integrated concurrence
are measures for different types of entanglement.

\section{System}

\begin{figure}[t]
\includegraphics[width=3.4cm]{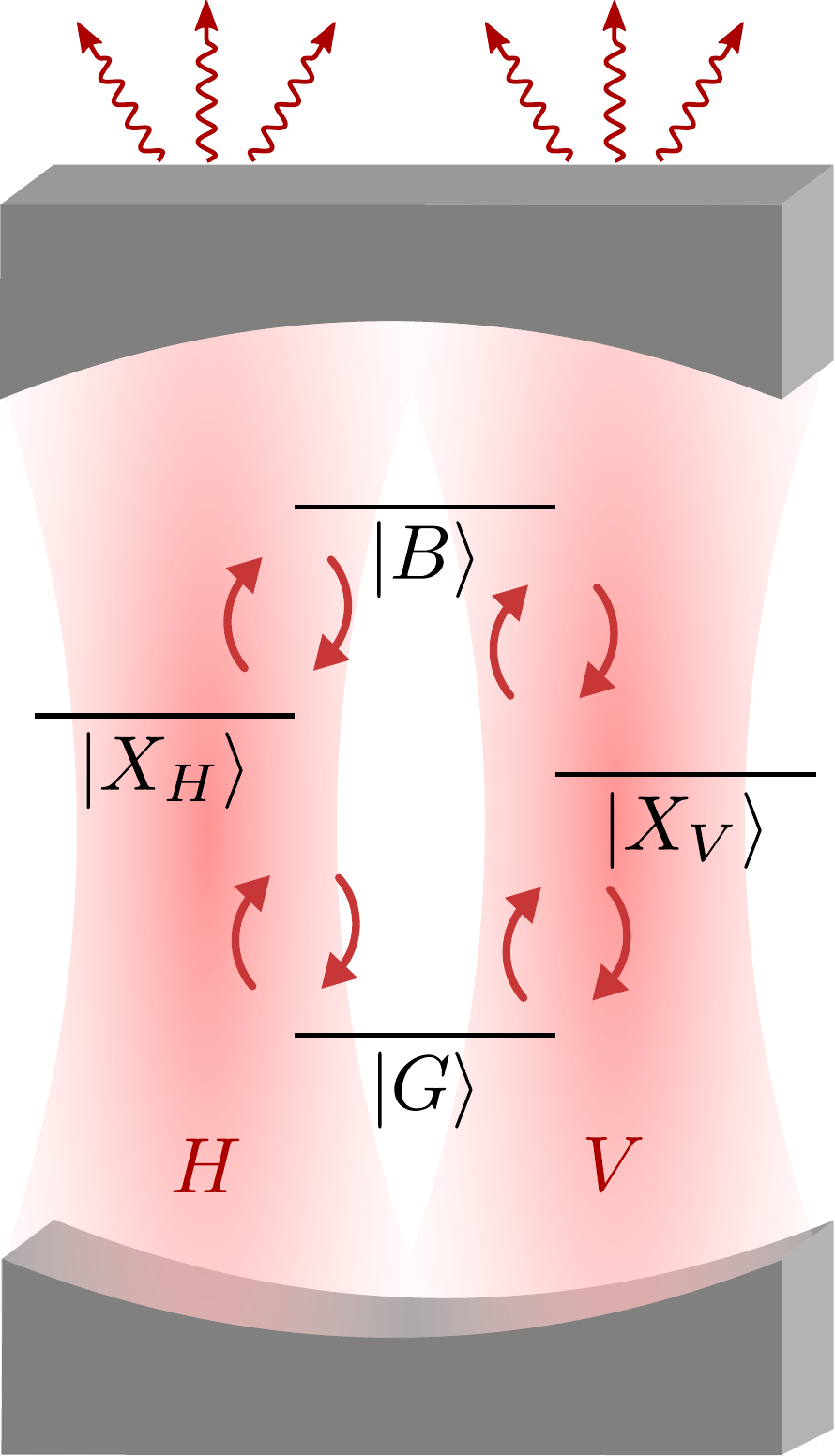}
\caption{\label{fig:sketch}
Sketch of the quantum dot states involved in the biexciton cascade 
inside a microcavity. The quantum dot is coupled to a microcavity 
with two orthogonally polarized cavity modes ($H$: horizontal, $V$: vertical).
Due to optical selection rules, the biexciton state ($|B\rangle$) can be deexcited
to one of two exciton states ($|X_H\rangle$ and $|X_V\rangle$) under the emission
of a $H$ or $V$ polarized cavity photon.
The exciton states can be further deexcited to the ground state ($|G\rangle$) of the 
dot by emitting a second photon. The reverse processes are also possible.
The red circular arrows indicate the dot-cavity coupling compatible with
the optical selection rules. Wavy arrows symbolize the photon losses
due to the imperfect cavity.
}
\end{figure}
\subsection{Hamiltonian}
We consider a quantum dot in a microcavity as depicted in Fig.~\ref{fig:sketch}.
We assume that the quantum dot is initialized at time $t=0$ in the biexciton 
state with an empty cavity, e.g., by incoherent excitation from the 
wetting layer or direct laser excitation.
The biexciton is coupled to two quantum states with an exciton in the 
quantum dot and a photon in the cavity. The two excitons are labeled
by $X_H$ and $X_V$ corresponding to the polarization of the cavity mode,
horizontal ($H$) or vertical ($V$), to which the respective transition 
is coupled. The excitonic states are also coupled to the ground state of the
dot with one more photon in the cavity. 
At the same time, the cavity is subject to losses and the
excitons in the dot interact with longitudinal acoustic (LA) phonons.
Radiative decay is assumed to be negligible compared with cavity losses,
which is a typical case \cite{reithmaier04}.

The dot-cavity Hamiltonian is given by \cite{Heinze2017}
\begin{align}
H_{dc}=&\;
\big(\hbar\bar{\omega}_X+\frac \delta 2\big)|X_H\rangle\langle X_H|
+\big(\hbar\bar{\omega}_X-\frac \delta 2\big)|X_V\rangle\langle X_V|
	\nn&
+\big(2\hbar\bar{\omega}_X-E_B\big)|B\rangle\langle B|
+\hbar\omega_H a^\dagger_Ha_H+\hbar \omega_V a^\dagger_Va_V\nn&
+\hbar g\Big[\big(|G\rangle\langle X_H| +|X_H\rangle\langle B|\big)a^\dagger_H
	\nn&
+\big(|G\rangle\langle X_V| - |X_V \rangle\langle B|\big) a^\dagger_V
	\nn& 
+ \big(|X_H\rangle\langle G| + |B\rangle\langle X_H|\big) a_H
	\nn&
+\big(|X_V\rangle\langle G| - |B \rangle\langle X_V|\big) a_V \Big],
\label{eq:Hamiltonian}
\end{align}
where $\hbar\bar{\omega}_X$ is the average exciton energy, $\delta$ is the
fine structure splitting and $E_B$ is the biexciton binding energy. 
The energies of the cavity modes are $\hbar\omega_H$ and $\hbar\omega_V$,
respectively, and $g$ is the dot-cavity coupling constant.
Here, we assume that the cavity modes are in resonance with the two-photon 
transition between the ground and the biexciton state
$\hbar\omega_H=\hbar\omega_V=\frac{2\hbar\bar{\omega}_X-E_B}2$.
$|G\rangle$ denotes the ground state of the dot, $|X_H\rangle$ and $|X_V\rangle$
are the exciton states, and $|B\rangle$ is the biexciton state.
Finally, $a^\dagger_i$ ($a_i$) is the creation (annihilation) operator
for a photon in the cavity mode $i=\{H,V\}$.

Cavity losses are taken into account via the Lindblad term
\begin{align}
\mathcal{L}_\textrm{cavity}[\rho]=\frac \kappa 2\sum_{i=H,V}
\big(2a_i\rho a^\dagger_i-a^\dagger_ia_i\rho-\rho a^\dagger_i a_i\big),
\label{eq:lossterm}
\end{align}
where $\kappa$ is the cavity loss rate.
The interaction between the dot and the LA phonons is described by the
Hamiltonian
\begin{align}
\hat{H}_{ph}=\hbar \! \sum_{\mathbf q}\omega_{\mathbf q}{b}^\dagger_{\mathbf q} {b}_{\mathbf q}
+\hbar \! \sum_{\mathbf q}\sum_{\nu} n^X_\nu
\big(\gamma_{\mathbf q}^X {b}^\dagger_{\mathbf q}
+\gamma_{\mathbf q}^{X*}{b}_{\mathbf q}\big)
|\nu\rangle\langle \nu| ,
\label{eq:Hosc}
\end{align}
where $b^\dagger_{\mathbf q}$ and $b_{\mathbf q}$ are the creation and 
annihilation operators for phonons with wave vectors $\mathbf q$ and
energies $\hbar\omega_{\mathbf q}$. $n^X_\nu=\{0,1,1,2\}$ 
is the number of excitons
in the dot state $\nu=\{G,X_H,X_V,B\}$ and $\gamma_{\mathbf q}^X$ 
is the exciton-phonon coupling constant.

\subsection{Concurrences}

The time-dependent concurrence which has been shown by Wootters \cite{Wootters1998}
to be a measure for the entanglement of formation is, for 
the two-photon state created in the biexciton cascade,
given by Eq.~\eqref{eq:C}.
Since the density matrix in this equation is evaluated with cavity operators [cf. Eq.~\eqref{eq:twophotondensmat}],
it immediately follows that $C(t)$ contains information about the entanglement
of formation of the two-photon state inside the cavity at any given time $t$.
Experimentally, the reduced two-photon density matrix is typically reconstructed via quantum state tomography, 
where polarization-resolved two-photon coincidence rates are measured which are proportional to the 
two-time correlation functions
\begin{align}
G^{(2)}_{ij,kl}(t,\tau)=\langle a_i^\dagger(t) a_j^\dagger(t+\tau) a_k(t+\tau) a_l(t)\rangle.
\label{eq:G2}
\end{align}
Here, $t$ is the time of the first click at a detector,  $\tau$ is
the delay time until the second photon is detected and $i,j,k,l\in\{H,V\}$.
Since in experiments one measures photons that have left the cavity,
the cavity operators in Eq.~\eqref{eq:G2} should in fact be
replaced by operators for the field modes outside the cavity.
However, considering the outcoupling of light out of the cavity to be a Markovian
process, the quantities measured outside the cavity are proportional to the ones
inside\cite{Kuhn:16}.
Therefore, a measurement of $G^{(2)}_{ij,kl}(t,\tau)$ outside the cavity can indeed be 
described by Eq.~\eqref{eq:G2}.
Finally, we note that in the present analysis we have assumed the radiative decay to
be negligible compared with the cavity losses and, consequently,
we only consider the case that photons are emitted via the cavity.
For a direct emission of photons into modes outside the cavity by radiative decay,
two-time correlation functions involving polarization operators instead of  photon operators
would have to be considered \cite{Troiani2006}.

Typically, corresponding experiments record data points over extended time intervals
$t_{0}\le t\le t_{0}+ \Delta t$ and $\tau_{0}\le\tau\le \tau_{0}+\Delta\tau$.
For the reconstruction of the unnormalized density matrix defined in Eq.~\eqref{eq:twophotondensmat}
the delay line between the two detectors measuring the coincidence is adjusted such that
the two intervals where the detectors are sensitive start simultaneously, i.e. $\tau_{0} = 0$.
Setting $t_{0}=0$ the measured signals are then proportional to
\begin{align}
&\overline{\overline G}^{(2)}_{ij,kl}(\Delta t,\Delta \tau)=
\notag\\
&\frac{1}{\Delta t\,\Delta\tau}
\int_{0}^{\Delta t} 
dt
\int_{0}^{\Delta \tau}
d\tau \;
\langle a_i^\dagger(t) a_j^\dagger(t+\tau) a_k(t+\tau) a_l(t)\rangle.
\label{eq:G2-av}
\end{align}
Thus, the result of an experimental reconstruction of the normalized two-photon density matrix is \cite{Troiani2006}
\begin{align}
    \rho^{R}_{j,l}(\Delta t,\Delta\tau) = 
\frac{\overline{\overline G}^{(2)}_{jj,ll}(\Delta t,\Delta\tau)}{\overline{\overline G}^{(2)}_{HH,HH}(\Delta t,\Delta\tau)+
\overline{\overline G}^{(2)}_{VV,VV}(\Delta t,\Delta\tau)}.
\end{align}

Associated with the reconstructed density matrix is the concurrence
\begin{align}
  C^{R}(\Delta t,\Delta\tau) =2|\rho^{R}_{H,V}(\Delta t,\Delta\tau)|.
\label{CR}
\end{align}
As low counting rates limit the experimental accuracy, common measurements are performed
using rather long intervals for the data collection. These experiments often
approach the limiting case $\Delta t\to\infty$ and
$\Delta \tau\to\infty$ such that in theory the corresponding concurrence 
becomes\cite{Jahnke2012,Heinze2017}
\begin{align}
  \overline{\overline{C}} = \lim_{
  \substack{
\Delta t\to\infty \\ \Delta \tau\to\infty 
}}
 C^{R}(\Delta t,\Delta\tau). 
\label{eq:dtC}
\end{align}
Due to the double-time averaging, the information about the
time evolution of the system is completely lost such that,
in contrast to the time dependent concurrence $C(t)$,
the double-time integrated concurrence $\overline{\overline{C}}$
does not reflect properties of the cavity photons at any given time
but rather describes the properties of the reconstructed density matrix in 
experiments.

While the case of extended measuring intervals for both $t$ and $\tau$
is probably the most often discussed situation, in the literature
also another limiting case has been considered \cite{carmele11,EdV},
where the concurrence is defined as
\begin{align}
  \overline{C} &= \lim_{
  \substack{
\Delta t\to\infty \\ \Delta \tau\to0 
}}
 C^{R}(\Delta t,\Delta\tau). 
\label{eq:stC}
\end{align}
Experimentally, the limit $\Delta \tau\to0 $ can be performed using time-windowing techniques 
\cite{StevensonPRL2008} which record signals over
different delay-time windows $\Delta\tau$ and extrapolate to $\Delta\tau = 0$.
More recently it has been shown experimentally that using time bins with a width of $4\,$ps
is sufficient to resolve the full $\tau$ dependence of the signal if the exciton fine-structure
splitting is on a scale of a few tens of \textmu eV\cite{Bounouar_2018}.
Henceforth, we refer to the concurrence 
$\overline{C}$ defined in Eq.~\eqref{eq:stC}
as the single-time integrated concurrence.
If the photon pairs are emitted from a dot-cavity system in
the steady state, $\overline{C}$ is equivalent to $C(t)$ since the two-photon density matrix
no longer depends on time in such a case.
However, when considering dynamical systems that are, e. g., driven by laser
pulses, this equivalence in general no longer holds.

It should be noted that, as for the time-dependent concurrence $C(t)$, also for the single-time integrated concurrence
$\overline{C}$ 
it is not necessary 
to evaluate the two-time correlation function defined in Eq.~\eqref{eq:G2} since
in the limit $\Delta\tau\to0$ only the unnormalized density matrix
given by Eq.~\eqref{eq:twophotondensmat} enters the expression in Eq.~\eqref{eq:stC}.
However, one should be aware that $\overline{C}$ is not the time average of the time-dependent concurrence $C(t)$.
Instead, the time average of the concurrence is given by
\begin{align}
\langle C\rangle_T &:= 
\frac{1}{T}\int_0^T dt\, C(t) 
 \notag\\
&=
\frac{1}{T}\int_0^T dt\, \frac{2|\rho_{HH,VV}(t)|}{\rho_{HH,HH}(t)+\rho_{VV,VV}(t)},
\label{eq:C_av}
\end{align}
where $T$ corresponds to the averaging time.

In the following we shall compare the time-dependent concurrence $C(t)$, which measures the
entanglement of formation of the two-photon system inside the cavity at a given point in time, with
the double- and single-time integrated concurrences $\overline{\overline{C}}$
and $\overline{C}$, respectively, obtained as a result of different quantum state reconstruction
strategies that involve data collection over extended time intervals.
Due to the one-to-one correspondence between two-time correlation functions
of photon operators inside and outside the cavity shown in Ref.~\onlinecite{Kuhn:16},
the comparison between $C(t)$, $\overline{C}$, and $\overline{\overline{C}}$ can also
be interpreted in terms of the photons recorded in the two detectors used in
coincidence measurement. Since the photons outside the cavity propagate with the 
speed of light they can be recorded at a given time in one of the detectors 
only when they have been emitted from the cavity
at a retarded time that matches the flight time between cavity and detector.
Thus, recording the correlation function
$G^{(2)}_{ij,kl}(t,\tau)$ for given values of $t$ and $\tau$ selects
photons according to their emission time from the cavity.
Note that for the photons inside the cavity there is no obvious relation
between their emission times from the dot since standing wave modes localized 
in the cavity are excited and thus these excitations contribute to the two-time correlation functions
as long as the photons stay in the cavity.

The above analysis suggests that a density matrix 
constructed from the two-time correlation function
for given values of $t$ and $\tau$,
i.e. $\rho_{j,l}(t,\tau) = \frac{ G^{(2)}_{jj,ll}( t,\tau)}{G^{(2)}_{HH,HH}( t,\tau)+G^{(2)}_{VV,VV}(t,\tau)}$, 
represents a measurement of photons in two detectors where the photons are selected according to their 
respective emission times from the cavity.
Then, $C(t,\tau)=2|\rho_{H,V}(t,\tau)|$ can be interpreted as a measure for
the entanglement of formation of the
so selected photons. The question now arises how to
interpret the quantity $C^{R}(\Delta t,\Delta\tau)$
defined in Eq.~\eqref{CR}, which is obtained by first
taking a coherent superposition of signal contributions
associated with different emission times and then
constructing the concurrence by  taking twice the absolute
value of the off-diagonal element of the normalized superposition.
At this point it is helpful to discuss in more detail
the emission process from the cavity.
If we were dealing with an ensemble of randomly 
distributed emission events where each emission has a 
sharply defined emission time $t_{e}$ and where
the uncertainty concerning the emission time
is described by a classical probability distribution for $t_{e}$,
then the concurrence of this ensemble would be obtained
by first evaluating the concurrence separately for each ensemble
member, which would be $C(t,\tau)=2|\rho_{H,V}(t,\tau)|$,
and then averaging over the emission times, i.e., $t$ and $\tau$.
The result would be an average concurrence similar to
$\langle C\rangle_T$ except that here the average would
be taken over $t$ and $\tau$.
However, in a full quantum description, the uncertainty concerning the emission times 
is not represented by a classical ensemble of events with sharp emission times.
Instead, states where a photon has been emitted are typically 
in a coherent superposition with states where no photon has been emitted. 
Thus, the emission is not point-like in time but is a process of finite duration,
so that $\overline{\overline G}^{(2)}_{ij,kl}(\Delta t,\Delta \tau)$ can be thought of
as filtering out the coherent superposition corresponding to those contributions where the emission 
times are restricted to $t$ and $\tau$ intervals of lengths $\Delta t$ and $\Delta \tau$, respectively.
This means that $C^{R}(\Delta t,\Delta\tau)$ represents the concurrence associated with
a reconstructed two-photon density matrix which is filtered with respect to
emission times within finite intervals.
From this point of view, $C(t)$ represents the concurrence of photons recorded in two detectors
that are simultaneously (i.e. $\tau=\Delta\tau=0$) emitted from the cavity at a given time $t$, while
$\overline{C}$ also describes the concurrence of simultaneously emitted photons but without 
resolving their common emission time.
Finally, $\overline{\overline{C}}$ is the concurrence associated with a two-photon density matrix 
where one neither resolves the emission time of the first recorded photon
nor the delay time $\tau$ between the photons.

\section{Time-dependent concurrence}

In the following, we first present an approximate analytic expression for the
time-dependent concurrence of the two-photon state generated by the
biexciton cascade in a dot-cavity system in the absence of dot-phonon
interaction. Subsequently, we compare the anayltic results with numerical
calculations of the time-dependent concurrence as well as 
the single-time integrated concurrence with and without dot-phonon interaction.

\subsection{Analytic results}

In order to discuss how the time-dependent concurrence $C(t)$ depends
on the parameters of the system, it is instructive to look
for an approximate analytic solution 
of the dynamics in the absence dot-phonon interaction.

First, note that only few states contribute to the biexciton cascade:
The general states of the system can be described by $|\nu,n_H,n_V\rangle$,
where $\nu\in\{G,X_H,X_V,B\}$ denotes the dot state and $n_H$ and 
$n_V$ are the numbers of horizontally and vertically polarized cavity photons, 
respectively. Without losses and under the assumption that the system is initially prepared 
in the biexciton state $|B,0,0\rangle$ and not driven externally, the number of
total excitations (number of excitons plus number of photons) in the system
is fixed to two for all times.
When accounting for losses via the Lindblad operator defined in Eq.~\eqref{eq:lossterm},
also states with excitation numbers smaller than two become occupied.
However, these states need not be considered for the subsequent dynamics since first, 
they do not contribute to the two-photon density matrix $\rho_{ij,kl}(t)$
defined in Eq.~\eqref{eq:twophotondensmat} and second, states with lower excitation numbers
do not couple back to states with higher excitation numbers.
Since there are no direct transitions between the different excitons
$X_H$ and $X_V$  or between horizontally and vertically polarized photons,
only five remaining states contribute, which we denote by
\begin{subequations}
\begin{align}
&|G_H\rangle:=|G,2,0\rangle,\\
&|G_V\rangle:=|G,0,2\rangle,\\
&|X_H\rangle:=|X_H,1,0\rangle,\\
&|X_V\rangle:=|X_V,0,1\rangle,\\
&|B\rangle:=|B,0,0\rangle.
\end{align}
\label{eq:basis5lvl}
\end{subequations}
In this basis, the Hamiltonian in Eq.~\eqref{eq:Hamiltonian}
takes the form
\begin{align}
H=\left(\begin{array}{ccccc}
0&0&\sqrt{2}\hbar g&0&0\\
0&0&0&\sqrt{2}\hbar g&0\\
\sqrt{2}\hbar g &0&\frac 12(E_B+\delta)&0&\hbar g\\
0&\sqrt{2}\hbar g&0&\frac 12(E_B-\delta)&-\hbar g\\
0&0&\hbar g&-\hbar g&0
\end{array}\right),
\label{eq:H5by5}
\end{align}
where the origin of the energy scale is shifted to the biexciton.

An analytic solution of the full five-level system is complicated and
has so far only been presented in the weak coupling limit
\cite{carmele11} ($g\ll\kappa$), where only one-way transitions 
along the paths $|B\rangle \to |X_H\rangle \to |G_H\rangle$ and 
$|B\rangle \to |X_V\rangle \to |G_V\rangle$ can occur because the 
photon losses are much faster than the time needed for the reexcitation of
higher-energetic dot states.
However, to fully benefit from the microcavity one is often interested in
strongly coupled dot-cavity systems\cite{Reitzenstein_highQ,reithmaier08,
Kasprzak13}
 where the condition $g\ll\kappa$ is not
met and other approaches are required. 

Here, we make use of the fact that in typical quantum dots the 
biexciton binding energy $E_B\sim 1-6$~meV 
defines the largest energy scale.
Strongly coupled dot-cavity systems typically have couplings on the order of
$\hbar g\sim 0.1$~meV while typical values for the fine structure splitting 
are in the range of $\delta \sim 0.01-0.1$~meV, so that a perturbative 
treatment in terms of the small parameters $\lambda:=\hbar g/(\frac 12 E_B)$
and $\delta/E_B$ is appropriate. 
For later reference we also define 
$\lambda_{H/V}:=\hbar g/[\frac 12 (E_B\pm\delta)]$.

\begin{figure}[t]
\includegraphics[width=0.99\linewidth]{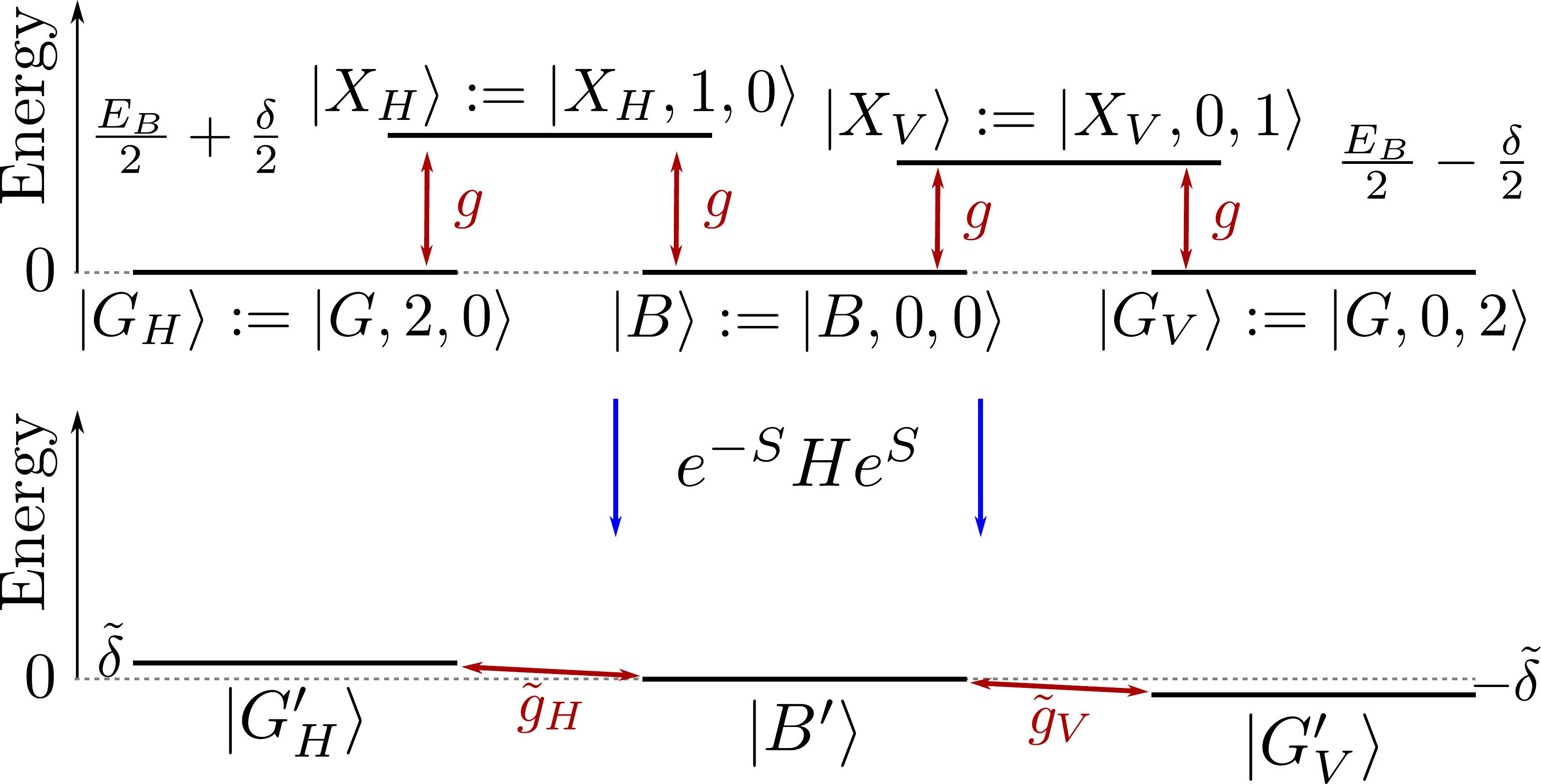}
\caption{\label{fig:sketch_3lvl}
Sketch of the block diagonalization reducing the dynamics in the 
five-level system
described in Eqs.~\eqref{eq:basis5lvl} to a three-level system 
with states defined in Eqs.~\eqref{eq:basis3lvl}.
}
\end{figure}

In the case considered here, where the cavity modes are in resonance 
with the two-photon transition to the biexciton state,
the comparatively large binding energy suppresses the occupation of the
exciton states. Thus, one can perform a perturbative block-diagonalization
\cite{Winkler,LossDivincenzo_SW}
(Schrieffer-Wolff transformation) and thereby remove the high-energy
states with one exciton and one photon from the dynamics, as sketched in
Fig.~\ref{fig:sketch_3lvl}.
To this end a unitary transform $e^{-S}He^{S}$ 
is applied to the Hamiltonian $H$, 
where $S=S^{(1)}+S^{(2)}+\dots$ is expanded in orders of the perturbing
Hamiltonian $H^{(1)}$ consisting of the off-diagonal elements of $H$ in
Eq.~\eqref{eq:H5by5}. The matrices $S^{(i)}$ are then obtained by the 
condition that the matrix elements between high- and low-energy states
in $e^{-S}He^{S}$ vanish up to order $\mathcal{O}\big( (H^{(1)})^{i+1}\big)$
(cf. Refs.~\onlinecite{Winkler} or \onlinecite{LossDivincenzo_SW} for explicit 
expressions for $S^{(i)}$). This transformation therefore
perturbatively eliminates the couplings between the low-energy 
ground- and biexciton-like states and the high-energy exciton-like states.
After decoupling the low-energy and high-energy states, the latter are
disregarded as they are irrelevant for the dynamics.

Up to second order in $\lambda$, the block-diagonalization yields the
effective Hamiltonian
\begin{align}
H'=\left(\begin{array}{ccc}
\tilde{\delta}&0&\hbar\tilde{g}_H \\
0&-\tilde{\delta}&\hbar\tilde{g}_V\\
\hbar\tilde{g}_H& \hbar \tilde{g}_V &0
\end{array}\right),
\end{align}
with $\tilde{\delta}=\hbar g(\lambda_V-\lambda_H)\approx \lambda^2 \delta$ and 
$\tilde{g}_{H/V}=\mp\sqrt{2}\lambda_{H/V} g$ in the basis 
\begin{subequations}
\begin{align}
|G'_H\rangle:=&\big(1-\lambda_H^2\big)|G_H\rangle-\sqrt{2}\lambda_H |X_H\rangle
-\frac 1{\sqrt{2}}\lambda_H^2|B\rangle, \\
|G'_V\rangle:=&\big(1-\lambda_V^2\big)|G_V\rangle-\sqrt{2}\lambda_V |X_V\rangle
+\frac 1{\sqrt{2}}\lambda_V^2|B\rangle,
\\
|B'\rangle:=&-\frac{\lambda_H^2}{\sqrt{2}}|G_H\rangle 
+\frac{\lambda_V^2}{\sqrt{2}}|G_V\rangle
-\lambda_H|X_H\rangle + \lambda_V|X_V\rangle \nn&
+\big(1-\frac 12(\lambda_H^2+\lambda_V^2)\big)|B\rangle.
\end{align}
\label{eq:basis3lvl}
\end{subequations}

Thus, perturbation theory in $\lambda$ allows one to reduce the five-level 
system of the biexciton cascade to an effective three-level system, where the
three levels have mostly the character of the ground state of the dot with
two horizonally or vertically polarized photons and the biexciton state.

In the three-level basis, the effective
coupling $\tilde{g}_{H/V}$ is reduced by a factor $\sim\lambda$ compared with
the coupling $g$ in the five-level system.
The effective splitting $2\tilde{\delta}$ between the states 
$|G'_H\rangle$ and $|G'_V\rangle$ is reduced even more 
compared with the fine structure splitting $\delta$ of the excitonic states in the
original five-level system because it only appears in second order in 
$\lambda$. 
When also the Lindblad terms are written in the basis described in 
Eq.~\eqref{eq:basis3lvl}, the biexciton-like state $|B'\rangle$ acquires 
the small loss rate $(\lambda_H^2+\lambda_V^2)\kappa$ and 
the loss rates for the states $|G'_H\rangle$ and $|G'_V\rangle$
become $2(1-\lambda_{H/V}^2)\kappa$, 
which are of the same order of magnitude as the rates $\kappa$
for the corresponding states $|G_H\rangle$ and $|G_V\rangle$ 
in the five-level system.

The central insight gained by this transformation is that, 
due to the renormalization of the coupling, the effective three-level
system can be in the weak coupling limit $|\tilde{g}_{H/V}| \ll \kappa$ even
when the orignal five-level system describing the biexciton cascade
is not $(g\sim \kappa)$, as is the case for typical parameters for 
dot-cavity systems \cite{Jahnke2012}.
Due to the weak coupling, the dynamics in the effective three-level system 
is easily understood:
The initial occupations of the biexciton state $|B'\rangle$ are transferred to
the ground states $|G'_H\rangle$ and $|G'_V\rangle$ and then decay
due to the losses before they can reexcite the biexciton state, 
yielding an essentially incoherent dynamics. 
An explicit calculation of the dynamics in the weakly coupled effective 
three-level system is presented in appendix \ref{app:3lvl}.
It is found that the occupation of the biexciton-like state $|B'\rangle$
decays exponentially with an effective rate 
\begin{align}
\kappa_{B}=(\lambda_H^2+\lambda_V^2)\left(\frac{4g^2}{\kappa} +\kappa\right),
\label{eq:kappaB}
\end{align}
where the term propotional to ${4g^2}/{\kappa}$ is due to the transitions
to the states $|G'_H\rangle$ and $|G'_V\rangle$ and the term proportional to
$\kappa$ originates from the losses due to the admixture of states with 
a nonvanishing number of photons to the state $|B'\rangle$.
The occupations and coherences 
$\rho_{G'_iG'_j}=\langle \big(|G'_i\rangle\langle G'_j|\big)\rangle$
between the states $|G'_H\rangle$ and $|G'_V\rangle$ are found to be
\begin{align}
&\rho_{G'_iG'_j}=
\frac{\tilde{g}_i\tilde{g}_j}{\kappa^2}\left(e^{-\kappa_B t} 
-e^{[-(2-\lambda_i^2-\lambda_j^2)\kappa
+i(\tilde\delta_i-\tilde\delta_j)/\hbar]t}\right),
\label{eq:solution3lvl}
\end{align}
with $\tilde\delta_{H/V}=\pm \tilde{\delta}$.

The long-time dynamics of $\rho_{G'_iG'_j}$ is determined by the same 
loss rate $\kappa_B$ as the biexciton-like state, whereas the 
initial increase from zero is governed by a term decaying with 
$\kappa\gg \kappa_B$. Because the renormalized splitting $\tilde{\delta}$
is very small compared with typical loss rates $\kappa$, possible oscillations
of $\rho_{G'_HG'_V}$ are overdamped. Furthermore, the second term 
in Eq.~\eqref{eq:solution3lvl} disappears already after a short
time $\sim \kappa^{-1}$. 
Neglecting $\tilde\delta$ in the second exponent in Eq.~\eqref{eq:solution3lvl}
yields the following simple analytic expression for the concurrence
\begin{align}
&C_\textrm{analytic}=\frac{2|\rho_{G'_HG'_V}|}{\rho_{G'_HG'_H}+\rho_{G'_VG'_V}}
\approx \frac{2|\tilde{g}_H\tilde{g}_V|}{\tilde{g}_H^2+\tilde{g}_V^2}
= \frac{E_B^2-\delta^2}{E_B^2+\delta^2}.
\label{eq:Canalyt}
\end{align}
First, we find that, although the density matrix
elements change in time, the analytic expression predicts that the concurrence
is constant in time. Furthermore, the concurrence depends only on the
biexciton binding energy and the fine structure splitting and is indepedent
of the dot-cavity coupling $g$ and the cavity loss rate $\kappa$.

Because all density matrix elements entering in the expression for 
the concurrence have virtually the same time dependence, integrating
the density matrix elements over the time $t$ yields the same result 
also for the single-time integrated concurrence 
\begin{align}
&\overline{C}_\textrm{analytic}=C_\textrm{analytic}\approx
\frac{E_B^2-\delta^2}{E_B^2+\delta^2}.
\label{eq:Canalyt2}
\end{align}
Thus, our analysis reveals that the single-time integrated concurrence
is the same as the concurrence at any point in time and
is therefore a measure of the entanglement of formation
for the two-photon state generated in the cavity by the biexciton cascade.

\subsection{Numerical results}

To check the validity of the analytic results for the concurrence,
we now present numerical calculations of the biexciton cascade 
described by the dot-cavity Hamiltonian in Eq.~\eqref{eq:Hamiltonian} and the 
loss term in Eq.~\eqref{eq:lossterm} in the five-level basis introduced in 
Eq.~\eqref{eq:basis5lvl}. Futhermore, we study the effects of phonons
due to the dot-phonon Hamiltonian in Eq.~\eqref{eq:Hosc}, 
which have been neglected in the derivation of the analytic results, 
using a numerically exact real-time path-integral method 
\cite{Makri_Theory,PI_realtime2011,PI_nonHamil2016,PI_cQED} 
described in detail in the supplement of 
Ref.~\onlinecite{PI_cQED}.

If not stated otherwise, we use the following parameters: dot-cavity coupling
constant $\hbar g=0.1$~meV, biexciton binding energy $E_B=1.5$~meV,
cavity loss rate $\kappa=0.25$~ps$^{-1}$, and fine structure splitting
$\delta=0.1$~meV.
Note that for these parameters ($g/\kappa\approx 0.6$)
the system is clearly not in the weak-coupling limit, so that 
conventional weak-coupling theories are not applicable.
For calculations involving the dot-phonon interaction,
we use parameters suitable for a 3 nm wide self-assembled InGaAs quantum dot
embedded in a GaAs matrix (cf. Ref.~\onlinecite{PI_cQED}).
Furthermore, the phonons are assumed to be initially in equilibrium at a
temperature $T=10$~K.

\begin{figure}[t]
\includegraphics{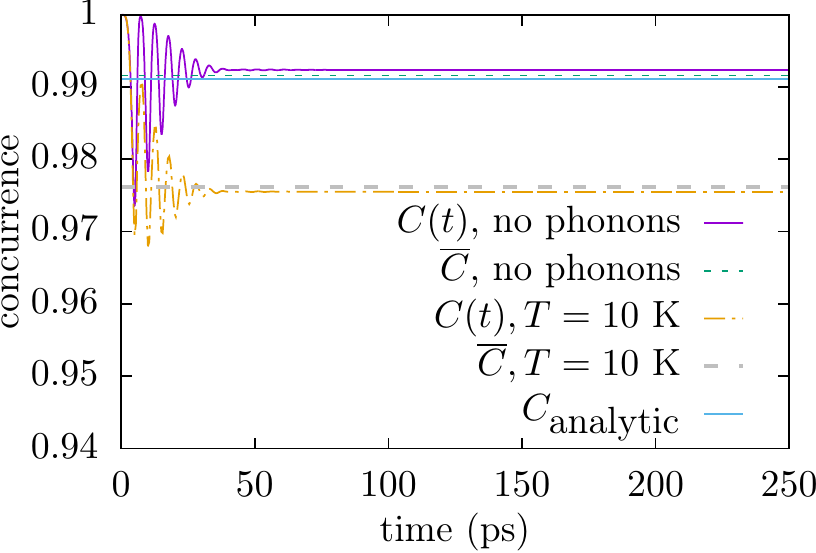}
\caption{Time-dependent concurrence $C(t)$ and single-time integrated
concurrence $\overline{C}$ calculated numerically without dot-phonon interaction 
and with phonons at a temperature $T = 10$~K compared with the
analytically obtained result $C_\textrm{analytic}$ for the phonon-free case.
\label{fig:timedepC}}
\end{figure}

Figure~\ref{fig:timedepC} depicts 
the time evolution of the time-dependent 
concurrence $C(t)$ and the single-time integrated concurrence $\overline{C}$
determined numerically as well as its analytic value according to Eq.~\eqref{eq:Canalyt2}.
In the absence of dot-phonon interaction,
the time-dependent concurrence indeed agrees well
with the constant analytic result as well as the single-time integrated 
concurrence after an initial phase of $\sim 40$~ps duration, as expected from 
the analytic results.
If phonons are taken into account, $C(t)$ and $\overline{C}$ still agree
well after this initial phase, but the stationary value for long times
is reduced.

\begin{figure}[t]
\includegraphics{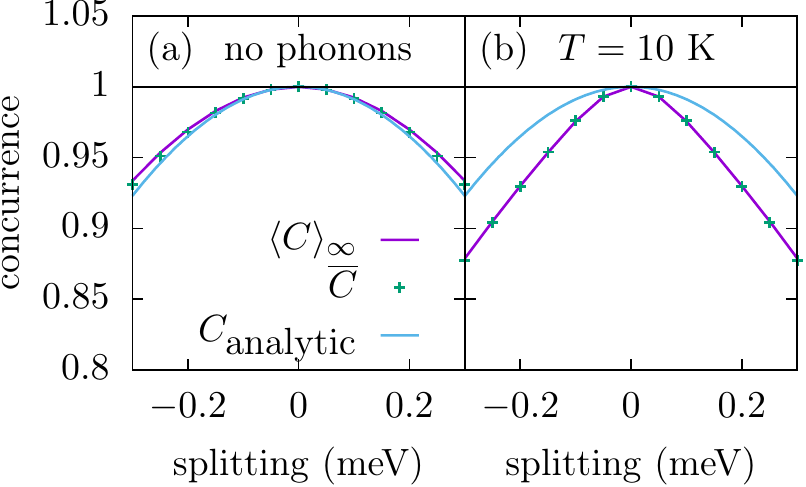}
\caption{Dependence of the time-averaged concurrence $\langle C\rangle_\infty$, 
and the single-time integrated concurrence $\overline{C}$
on the fine structure splitting $\delta$,
calculated (a) without dot-phonon interaction and (b) with phonons 
at a temperature $T=10$~K. For comparison
the analytically obtained result $C_\textrm{analytic}$ for the phonon-free case is displayed in (a) and (b).
\label{fig:Cstivsdelta}}
\end{figure}

The dependence of the time-averaged concurrence $\langle C\rangle_\infty$
and the single-time integrated concurrence $\overline{C}$ on the fine structure splitting $\delta$
is shown in Fig.~\ref{fig:Cstivsdelta}.
Here, we consider $\langle C\rangle_\infty$ since it condenses the information contained in $C(t)$ 
into a single number that can be compared with $\overline{C}$.
The averaging time $T=1000$~ps for $\langle C\rangle_\infty$ is chosen such that it is much larger 
compared with all other timescales in the system.
For calculations not accounting for the dot-phonon interaction [Fig.~\ref{fig:Cstivsdelta}(a)],
the time-averaged concurrence and the single-time integrated concurrence 
are in good agreement with the analytic results for the whole range of fine structure splittings.
When phonons are taken into account [Fig.~\ref{fig:Cstivsdelta}(b)], 
both definitions of the concurrence still coincide but yield, in general, significantly lower values than 
the concurrence obtained by neglecting the dot-phonon interaction.
However, at vanishing fine structure splitting $\delta=0$, the concurrence
remains one even in the presence of phonons as predicted in 
previous studies\cite{carmele11}. This is due to the fact
that both paths of the biexciton cascade are completely symmetric in this case.
The resulting absence of which-way information makes it possible to
get a completely entangled state for all times.

It is worth noting that the value of the concurrence remains close to
one even for the relatively large fine structure splitting of $\delta=0.1$~meV. 
As will be discussed in Sec.~\ref{sec:double-time-integrated-concurrence}, the
dependence of the double-time integrated concurrence on $\delta$ turns out to be
completely different.
The near independence on $\delta$ of the single-time integrated concurrence is easily understood by
the argument of Stevenson \emph{et al.}~\cite{StevensonPRL2008} discussed in the introduction according to
which there is a high probability for finding the maximally entangled state 
$|\Psi\rangle=\frac 1{\sqrt{2}}\big( |HH\rangle +e^{i\tau\delta/\hbar}|VV\rangle\big)$.
Thus, the time-dependent as well as the single-time integrated concurrence have high
values because even for finite $\delta$ the system is at any point in time
close to a maximally entangled state. This high degree of entanglement can, however,
not be uncovered when the system state is reconstructed by collecting data points over an extended
$\tau$ interval, as is done in the double-time integrated concurrence, because of destructive interference. 
The remaining observed weak decrease of the single-time integrated concurrence with rising $\delta$ reflects 
the deviation of the actual system state from the idealized pure state $|\Psi\rangle$.

\begin{figure}[t]
\includegraphics{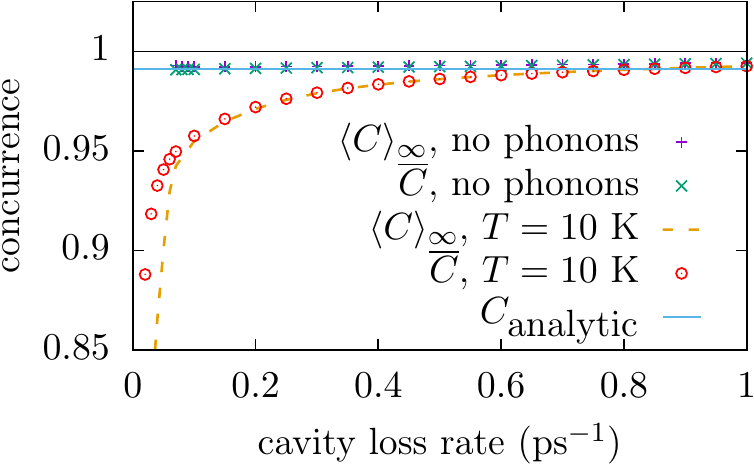}
\caption{Dependence of the time-averaged concurrence $\langle C\rangle_\infty$ and
the single-time integrated concurrence $\overline{C}$ with and without dot-phonon
interaction on the cavity loss rate $\kappa$ using a fine structure 
splitting $\delta = 0.1$~meV.
Also shown is the analytically obtained result $C_\textrm{analytic}$.
\label{fig:Cstivskappa}}
\end{figure}

Finally, the dependence of $\langle C\rangle_\infty$  and the
single-time integrated concurrence $\overline{C}$ on the
cavity loss rate $\kappa$ is depicted in Fig.~\ref{fig:Cstivskappa}
for a fine structure splitting $\delta = 0.1$~meV.
Again, in absence of dot-phonon interaction,
the time-averaged concurrence as well as the single-time integrated concurrence 
coincide with the analytic result, which is independent of $\kappa$.
When the interaction between the quantum dot and the phonons is accounted for,
it is found that the concurrence increases monotonically with increasing 
loss rate. At $\kappa=0$, the concurrence becomes zero (not shown)
and for large loss rates, the concurrence approaches the same value as obtained
when the dot-phonon interaction is disregarded. 
This can be explained by the fact that the dot-phonon interaction 
enables phonon-assisted processes. First of all, given enough time, 
phonon emission and absorption leads to a thermal occupation of 
energy eigenstates. Secondly, transitions involving the exciton states 
$|X_H,1,0\rangle$ and $|X_V,0,1\rangle$, which are otherwise off-resonant,
are enabled by the absorption of phonons with energies close to $E_B/2$.
For typical biexciton binding energies in the range of a few meV 
this energy is close to the maximum of typical phonon spectral density
\cite{PI_phonon-assisted_biexc_prep}, 
so that the phonon-assisted transitions through excitonic states 
are particularly efficient. In any case, the coherences between the states
$|G,2,0\rangle$ and $|G,0,2\rangle$ are strongly reduced by phonon effects, 
which in turn reduce the concurrence $C(t)$. 

However, the phonon-induced loss of coherence requires a finite amount
of time and therefore competes with the cavity losses. Note that the latter
leads to a uniform decrease of the coherences as well as the occupations of the 
two-photon state, which appear in the numerator and the denominator in the
definition of the concurrence $C(t)$ in Eq.~\eqref{eq:C}, respectively.
Therefore, the cavity losses do not directly affect the concurrence $C(t)$,
which is also the reason why the analytic expression $C_\textrm{analytic}$
in Eq.~\eqref{eq:Canalyt} does not depend on $\kappa$.
In contrast, phonons only marginally affect the occupations but they can
strongly reduce the coherences, which results in a reduced concurrence $C(t)$.
Thus, if the cavity loss rate $\kappa$ is small, phonons can suppress the 
degree of entanglement measured by the concurrence $C(t)$. For large 
$\kappa$, the occupations can decay faster than the time needed for
phonon-induced decoherence, so that for very large
$\kappa$ the phonon-free situation is recovered.

\begin{figure}[t]
\includegraphics{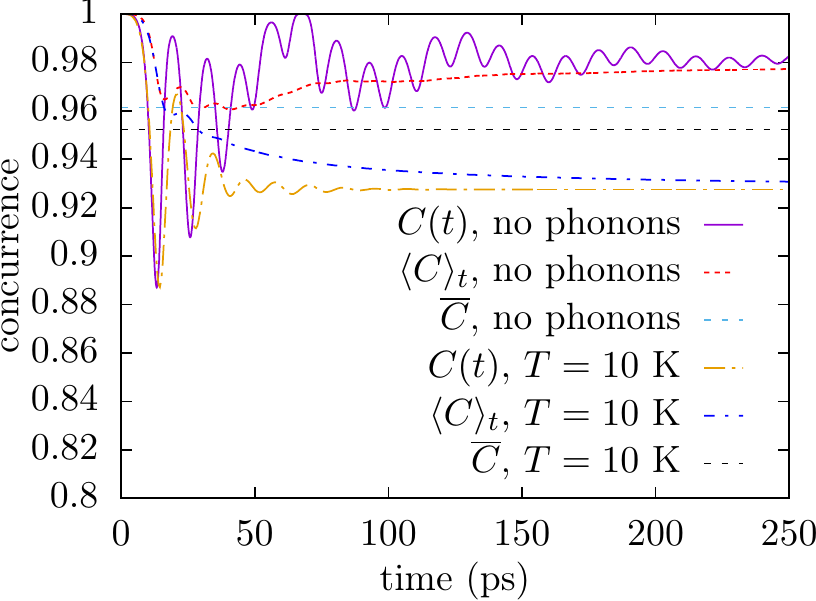}
\caption{Time-dependent concurrence $C(t)$ and time-averaged concurrence $\langle C\rangle_t$ 
as a function of time plotted together with the value of  the single-time integrated concurrence $\overline{C}$
for a quantum dot with a small biexciton binding energy $E_B = 0.3$~meV with and without dot-phonon interaction.
\label{fig:C_small_EB}}
\end{figure}

To summarize, our numerical calculations of the concurrence in the biexciton 
cascade in the absence of dot-phonon interaction confirm the validity of 
the analytic expression for the concurrence in Eq.~\eqref{eq:Canalyt}
for a large range of fine structure splittings and cavity loss rates.
This supports the core idea of our analytic approach 
that the biexciton cascade can be discussed in terms
of an effective three-level system that, when the biexciton binding energy
is large enough, is in the weak-coupling limit ($\tilde{g}\ll \kappa$)
even if the original system is not ($g\sim \kappa$). As a consequence 
of the weak coupling in the effective three-level system,
the dynamics of the relevant two-photon density matrix elements
is exponentially damped rather than oscillatory, so that an integration over time
does not lead to a cancellation of coherences.
For this reason, the single-time integrated concurrence $\overline{C}$
agrees very well with the typical value of the time-dependent concurrence $C(t)$
and its average value given by $\langle C\rangle_\infty$ and is therefore a measure 
for the entanglement of formation of the two-photon state.
Numerically exact path integral calculations reveal that the relation 
$\overline{C}\approx \langle C\rangle_\infty$ still holds when the dot-phonon interaction
is accounted for.

However, the quantitative agreement between $\overline{C}$ and the time-averaged concurrence $\langle C\rangle_\infty$
can also be brought to its limits:
For systems with small or vanishing biexciton binding energy, $\lambda = \hbar g/(\frac{1}{2}E_B)$ is
no longer a small parameter, which 
has consequences for the time-dependent concurrence as can be seen in Fig.~\ref{fig:C_small_EB}.
For a small biexciton binding energy, the concurrence $C(t)$ shows pronounced oscillations that persist
for more than $100$~ps if the influence of phonons is disregarded.
When accounting for the dot-phonon interaction the oscillations are damped down much more quickly.
These oscillations can be traced back to a coherent Rabi dynamics between the ground and the biexciton state
which causes an oscillatory behavior for both the numerator and the denominator in the expression for $C(t)$
given by Eq.~\eqref{eq:C}.
If one compares in such a case the long-time limit of the  time-averaged concurrence $\langle C\rangle_\infty$,
which  coincides reasonably well with $C(t)$ for long times,
with the single-time integrated concurrence $\overline{C}$,
where numerator and denominator are separately averaged over time, it is found that these quantities
noticeably deviate.

\section{Double-time integrated concurrence}
\label{sec:double-time-integrated-concurrence}

Having discussed the time-dependent and single-time integrated concurrence,
we now move on to the double-time integrated concurrence.
To this end, we derive an analytic expression for the double-time integrated 
concurrence for the biexciton cascade in absence of dot-phonon interaction
and subsequently compare it with numerical results.

\subsection{Analytic results}

The calculation of the double-time integrated concurrence 
$\overline{\overline{C}}$ as defined in Eq.~\eqref{eq:dtC}
requires the knowledge of the two-time correlation function
$G^{(2)}_{ij,kl}(t,\tau)$, which can be obtained in the 
Heisenberg picture by
\begin{align}
&G^{(2)}_{ij,kl}(t,\tau)=\textrm{Tr}\big[ a^\dagger_i(t)
a^\dagger_j(t+\tau)a_k(t+\tau)a_l(t)\hat\rho(0)\big],
\end{align}
where $\hat\rho(0)$ is the initial density matrix.
Introducing the time evolution operator $U(t)$ and rearranging terms
yields
\begin{align}
&G^{(2)}_{ij,kl}(t,\tau)=\textrm{Tr}\big[a_j^\dagger a_k U(\tau)
[a_lU(t)\hat\rho(0)U(-t)a_i^\dagger]U(-\tau)\big].
\end{align}
To obtain the double-time integrated concurrence, we first integrate
over the time $t$ and define
\begin{subequations}
\begin{align}
&\overline{G}^{(2)}_{ij,kl}(\tau)
=\int\limits_0^\infty dt\,G^{(2)}_{ij,kl}(t,\tau)=
\textrm{Tr}\big[a_j^\dagger a_k U(\tau)
\tilde{\rho}(0)U(-\tau)\big]
\nn&=
\textrm{Tr}\big[a_j^\dagger(\tau) a_k(\tau) \tilde{\rho}(0)\big]
=\langle a_j^\dagger a_k(\tau)\rangle_{\tilde\rho},
\label{eq:derivG2-a}
\end{align}
with
\begin{align}
&\tilde{\rho}(0)=\int\limits_0^\infty dt\,a_lU(t)\hat\rho(0)U(-t)a_i^\dagger
=a_l \bigg[\int\limits_0^\infty dt\, \hat\rho(t) \bigg]a_i^\dagger.
\end{align}
\label{eq:derivG2}
\end{subequations}

Thus, $\overline{G}^{(2)}_{ij,kl}(\tau)$ can be calculated like the
average of the operator $a_j^\dagger a_k$ at time $\tau$ evaluated 
with a generalized (possibly non-Hermitian) density matrix $\tilde\rho$.
The initial value of $\tilde\rho$ can be obtained from
the dynamics of the density matrix elements that have been calculated
analytically in the last section. The matrix elements have to be integrated 
over the time $t$ and the photon operators $a_l$ and $a_i^\dagger$ have to be 
applied from the left and from the right, respectively.
Note that to derive Eqs.~\eqref{eq:derivG2} we have assumed a time evolution
given by a unitary operator $U(t)$. In practice, we take into account a 
Lindblad term in the equations of motion for the description of cavity losses 
due to the coupling of the light modes within the cavity with 
a continuum of light modes outside the cavity.
Such non-Hamitonian terms give rise to a dynamics that is, in general,
 not described by a unitary time evolution. 
However, when the light modes outside the cavity 
are included in the description, the time evolution of the total system
can again be represented by a unitary time evolution operator $U(t)$. 
Calculating the trace in Eq.~\eqref{eq:derivG2-a} over the field modes outside
of the cavity and applying the usual Markovian approximation for the derivation
of the Lindblad equations it is straightforward to 
show that the abovementioned prescription for the calculation of
$\overline{G}^{(2)}_{ij,kl}(\tau)$ identically transfers to systems 
with Lindblad terms such as cavity losses.

As before, it is easy to see that in order to calculate the concurrence, 
one only needs to account for matrix elements of $\hat\rho(t)$ 
involving the five states defined in Eqs.~\eqref{eq:basis5lvl} with
exactly two excitations. The application of the operators
$a_l$ and $a_i^\dagger$ reduces the number of photons and thereby the 
number of excitations by one. Furthermore, the biexciton state without photons
$|B,0,0\rangle$ is removed by the action of a photon destruction operator, 
leaving only the four relevant states 
\begin{subequations}
\begin{align}
\tilde{G}_H=&|G,1,0\rangle,\\
\tilde{X}_H=&|X_H,0,0\rangle,\\
\tilde{X}_V=&|X_V,0,0\rangle,\\
\tilde{G}_V=&|G,0,1\rangle
\end{align}
\label{eq:basis4lvl}
\end{subequations}
that have to be accounted for in the calculation of $\tilde\rho(\tau)$. 
Restricted to this basis, the dot-cavity Hamiltonian reads
\begin{align}
&H=\left(\begin{array}{cccc}
0&\hbar g & 0 & 0\\
\hbar g& \frac 12(E_B+\delta)&0&0\\
0&0&\frac 12(E_B-\delta)&\hbar g \\
0&0&\hbar g& 0
\end{array}\right).
\end{align}
This Hamiltonian represents a system of two decoupled two-level systems,
which is diagonalized by the eigenstates
\begin{subequations}
\begin{align}
\tilde{G}'_i=&\cos(\lambda_i) |\tilde{G}_i\rangle
-\sin(\lambda_i)|\tilde X_i\rangle,\\
\tilde{X}'_i=&\sin(\lambda_i) |\tilde{G}_i\rangle
+\cos(\lambda_i)|\tilde X_i\rangle
\end{align}
\label{eq:diagBasis2lvl}
\end{subequations}
with $i\in\{H,V\}$.
In order to get more transparent expressions, we again focus on terms 
up to second order in $\lambda$ and approximate
$\cos(\lambda_{H/V})\approx 1-\frac 12\lambda_{H/V}^2$ as well as
$\sin(\lambda_{H/V})\approx \lambda_{H/V}$. Then, the  energy eigenvalues are
\begin{subequations}
\begin{align}
E_{\tilde G'_i}=&-\lambda_i^2 \frac 12(E_B+\delta_i),\\
E_{\tilde X'_i}=&(1+\lambda_i^2) \frac 12(E_B+\delta_i)
\end{align}
\end{subequations}
with $\delta_{H/V} = \pm\delta$.

\begin{figure}[t]
\includegraphics[width=0.99\linewidth]{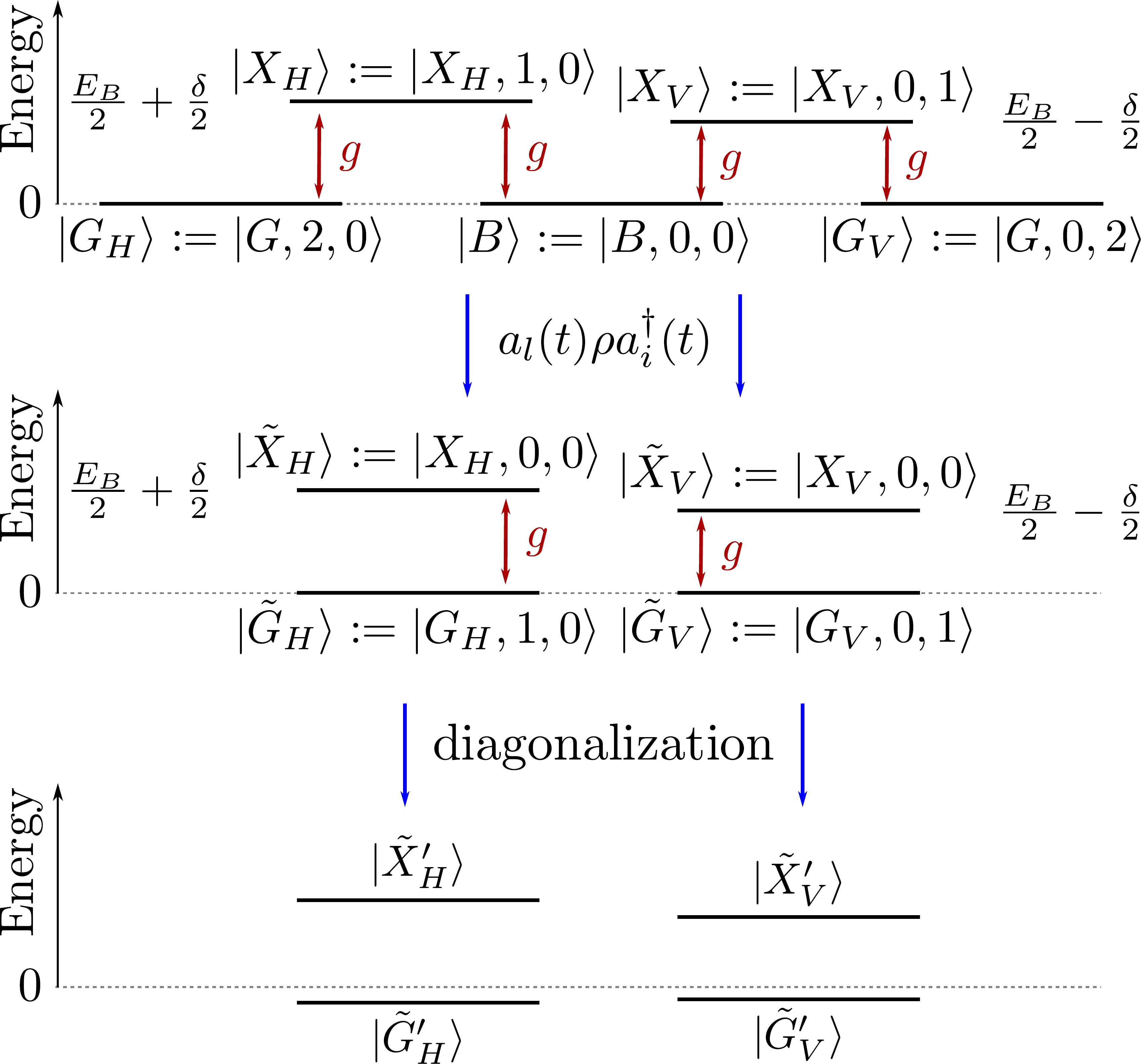}
\caption{\label{fig:sketch_4lvl}
Sketch of the four-level system for the description of the $\tau$-dependence 
of the two-time correlation function $G^{(2)}_{ij,kl}(t,\tau)$. The
application of $a_l$ from the left and $a_i^\dagger$ from the right on 
the density matrix $\rho$ with non-zero elements only for the five states 
depicted in the top panel results in a generalized density matrix
with non-zero elements only for the four states in the middle panel.
The resulting four-level system can be decomposed into two decoupled
two-level systems and diagonalized analytically.
}
\end{figure}

Transforming also the Lindblad term into the basis of the states in 
Eqs.~\eqref{eq:diagBasis2lvl}, we obtain the equations of motion
\begin{subequations}
\begin{align}
\ddtau \tilde{\rho}_{\tilde{G}'_i\tilde{G}'_j}=&
\big[\frac i\hbar(E_{\tilde{G}'_i}-E_{\tilde G'_j})
-\kappa\big(1-\frac 12(\lambda_i^2+\lambda_j^2)\big)\big]
\tilde{\rho}_{\tilde{G}'_i\tilde{G}'_j} \nn&
-\frac 12 \kappa \big(\lambda_i  \tilde{\rho}_{\tilde{X}'_i\tilde{G}'_j}
+\lambda_j  \tilde{\rho}_{\tilde{G}'_i\tilde{X}'_j}\big),\\
\ddtau \tilde\rho_{\tilde{G}'_i\tilde{X}'_j}=&
\big[\frac i\hbar(E_{\tilde{G}'_i}-E_{\tilde X'_j})
-\frac 12\kappa\big(1-\lambda_i^2+\lambda_j^2)\big)\big]
\tilde{\rho}_{\tilde{G}'_i\tilde{X}'_j} \nn&
-\frac 12 \kappa \big(\lambda_i  \tilde{\rho}_{\tilde{X}'_i\tilde{X}'_j}
+\lambda_j  \tilde{\rho}_{\tilde{G}'_i\tilde{G}'_j}\big),\\
\ddtau \tilde\rho_{\tilde{X}'_i\tilde{X}'_j}=&
\big[\frac i\hbar(E_{\tilde{X}'_i}-E_{\tilde X'_j})
-\frac 12\kappa\big(\lambda_i^2+\lambda_j^2)\big)\big]
\tilde{\rho}_{\tilde{X}'_i\tilde{X}'_j} \nn&
-\frac 12 \kappa \big(\lambda_i  \tilde{\rho}_{\tilde{G}'_i\tilde{X}'_j}
+\lambda_j  \tilde{\rho}_{\tilde{X}'_i\tilde{G}'_j}\big).
\end{align}
\label{eq:eomtilderho}
\end{subequations}
Note that the cross terms introduced by the losses are of minor importance
since the slowly changing occupations $\tilde\rho_{\tilde G'_i\tilde G'_j}$ and
$\tilde\rho_{\tilde X'_i\tilde X'_j}$ are off-resonantly driven by 
the fast oscillating (with frequency $\sim \frac 12 {E_B}/\hbar$) coherences 
$\tilde\rho_{\tilde G'_i \tilde X'_j}$ and vice versa.
This allows us to neglect these cross terms in the following.
Then, the solutions of Eqs.~\eqref{eq:eomtilderho} are damped oscillations.

Here, we are interested in the delay-time-integrated matrices
\begin{align}
&\overline{\overline{\rho}}_{ij}=\int\limits_0^\infty d\tau\,
\tilde{\rho}_{ij}(\tau),
\end{align}
which can be expressed by
\begin{subequations}
\begin{align}
\overline{\overline{\rho}}_{\tilde{G}'_i\tilde G'_j}=&
\frac{1}{\kappa \big(1-\frac 12(\lambda_i^2+\lambda_j^2)\big)
-\frac i\hbar \big(E_{\tilde G'_i} -E_{\tilde G'_j}\big)}
\tilde{\rho}_{\tilde{G}'_i\tilde G'_j}(0),\\
\overline{\overline{\rho}}_{\tilde{G}'_i\tilde X'_j}=&
\frac{1}{\frac 12\kappa \big(1-\lambda_i^2+\lambda_j^2\big)
-\frac i\hbar \big(E_{\tilde G'_i} -E_{\tilde X'_j}\big)}
\tilde{\rho}_{\tilde{G}'_i\tilde X'_j}(0),\\
\overline{\overline{\rho}}_{\tilde{X}'_i\tilde X'_j}=&
\frac{1}{\frac 12\kappa \big(\lambda_i^2+\lambda_j^2\big)
-\frac i\hbar \big(E_{\tilde X'_i} -E_{\tilde X'_j}\big)}
\tilde{\rho}_{\tilde{X}'_i\tilde X'_j}(0).
\end{align}
\label{eq:oorho}
\end{subequations}
The final steps to obtain the double-time integrated concurrence are a number of
basis transformations of the initial values:
First, we have to transform the analytic result
for the single-time averaged density matrix 
in the effective three-level system 
[basis: $G'_i, B'$ in Eqs.~\eqref{eq:basis3lvl}] back into the 
original five-level system [basis: $G_i, X_i,B$ in Eqs.~\eqref{eq:basis5lvl}],
then we have to apply the photon annihilation operators 
[new basis: $\tilde G_i, \tilde X_i$ in Eqs.~\eqref{eq:basis4lvl}] 
and transform the result to the diagonal basis spanned
by the states $\tilde{G}'_i$ and $\tilde X'_i$ defined in 
Eqs.~\eqref{eq:diagBasis2lvl} to obtain the initial 
values for the effective density matrix $\tilde\rho_{ij}(0)$.
With these initial values, Eqs.~\eqref{eq:oorho} are evaluated and
the result is transformed back to the basis spanned by $\tilde G_i$ and
$\tilde X_i$.

Keeping only second-order terms, we find the initial values:
\begin{subequations}
\begin{align}
\tilde\rho_{\tilde G'_i\tilde G'_j}(0)\approx&
\tilde\rho_{\tilde G_i\tilde G_j}(0)=
4 (2\delta_{ij}-1) \frac{g^2}{\kappa^2}\frac{\lambda_i\lambda_j}{\kappa_B},\\
\tilde\rho_{\tilde G'_i\tilde X'_j}(0)\approx&
\tilde\rho_{\tilde G_i\tilde X_j}(0)=
2i (2\delta_{ij}-1) \frac{g}{\kappa}\frac{\lambda_i\lambda_j}{\kappa_B},\\
\tilde\rho_{\tilde X'_i\tilde X'_j}(0)\approx&
\tilde\rho_{\tilde X_i\tilde X_j}(0)=
(2\delta_{ij}-1) \frac{\lambda_i\lambda_j}{\kappa_B}.
\end{align}
\end{subequations}
The double-time integrated density matrix elements entering the concurrence
are 
\begin{align}
\overline{\overline{\rho}}_{\tilde G_i \tilde G_j}\approx&
\frac{1-\frac 12(\lambda_i^2+\lambda_j^2)}{\kappa\big(
1-\frac 12(\lambda_i^2+\lambda_j^2)\big)}\tilde\rho_{\tilde G_i\tilde G_j}(0) 
\nn& 
+\frac{\lambda_i\lambda_j}{\frac 12\kappa (\lambda_i^2+\lambda_j^2)
-\frac i\hbar(E_{\tilde X'_i}-E_{\tilde X'_j})} 
\tilde\rho_{\tilde X_i\tilde X_j}(0)
\label{double-int-unnormalized}
\end{align}
because the coherences $\tilde\rho_{\tilde G'_i\tilde X'_j}$ lead to contributions
of the order of $\mathcal{O}(\lambda^3)$.
Thus, the double-time integrated density matrix has two contributions, one
from direct transitions through the low-energy eigenstates 
$\tilde G'_i$ and one from transitions through the high-energy exciton-like
eigenstates $\tilde X'_i$. On the one hand, the contributions from the 
occupations $\tilde\rho_{\tilde X'_i\tilde X'_j}$ are suppressed by a factor 
$\sim \lambda^2$ because the projection of $\tilde X'_i$ on $\tilde G_i$
is $\sim \lambda$. On the other hand, the losses for the exciton-like states
$\tilde X'_i$ are smaller by a factor $\sim\lambda^2$, so that the time integral
yields a larger contribution. All in all, the relative strength of the 
contributions from transitions through $\tilde G'_i$ and $\tilde X'_i$
are determined by the factor $4g^2/\kappa^2$.

It is also interesting that the occupations of the exciton-like states
 $\tilde X'_i$ stem from the projection of the occupations of 
biexciton-like eigenstate $B'$ onto $|X_H,1,0\rangle$ and $|X_V,0,1\rangle$
at the time of the loss of the first photon, i.e., when the first 
photon annihilation operator is applied. 
In contrast, the contributions through $\tilde G'_i$
have their origin in the occupations of the ground-states $|G,2,0\rangle$ 
and $|G,0,2\rangle$. This suggests that the latter can be interpreted as
a two-photon process in the sense that two excitations are transferred
from the quantum dot to the cavity before the first photon is emitted from 
the cavity, while the former corresponds to a one-photon process.
This interpretation is corroborated by the fact that the factor 
$4g^2/\kappa^2$ is identical to the ratio $\kappa_{2P}/\kappa_{1P}$
between cavity-assisted two- and one-photon emission processes discussed in
Ref.~\onlinecite{delValle_twoPhoton} in the context of the generation of
highly polarized (nonentangled) photon pairs.

Using the analytic expressions for the double-time integrated density matrix
$\overline{\overline{\rho}}$, the double-time integrated concurrence is found
to be
\begin{align}
\overline{\overline{C}}=&
\frac{2|\overline{\overline{\rho}}_{G_HG_V}|}
{\overline{\overline{\rho}}_{G_HG_H}+\overline{\overline{\rho}}_{G_VG_V}}\nn=&
C_\textrm{analytic}\frac{\Big|4\frac{g^2}{\kappa^2}+
C_\textrm{analytic}\frac{1}{1-i p}\Big|}
{4\frac{g^2}{\kappa^2}+1}
\end{align}
with
\begin{align}
&p:= \frac{E_{\tilde X'_H}-E_{\tilde X'_V}}
{\hbar\kappa \frac 12(\lambda_H^2+\lambda_V^2)}.
\end{align}
Keeping only the lowest-order terms in $\lambda$ and $\frac{\delta}{E_B}$ 
in the numerator and in the denominator, we can further simplify this 
result to
\begin{align}
&p\approx \frac{\delta}{\hbar \kappa \lambda^2}
=\frac{\delta E_B^2}{4\hbar^{3}\kappa g^2}.
\label{eq:ploworder}
\end{align}
Note that the contribution of $\frac 1{1-ip}$ to the concurrence
becomes insignificant for $\delta \gg \hbar\kappa\lambda^2$.
Therefore, this term only contributes
for small splittings $\delta$, for which the value of the 
concurrence $C_\textrm{analytic}$ is nearly one
and the double-time integrated concurrence is well described by
\begin{align}
\overline{\overline{C}}_\textrm{analytic}=&
C_\textrm{analytic}\frac{\sqrt{\Big(4\frac{g^2}{\kappa^2}+
\frac{1}{1+p^2}\Big)^2
+\Big(\frac{p}{1+p^2}\Big)^2}
}{4\frac{g^2}{\kappa^2}+1}
\label{eq:ooCanalyt}
\end{align}
with $p$ from Eq.~\eqref{eq:ploworder}.

\subsection{Numerical results}

\begin{figure}[t]
\includegraphics{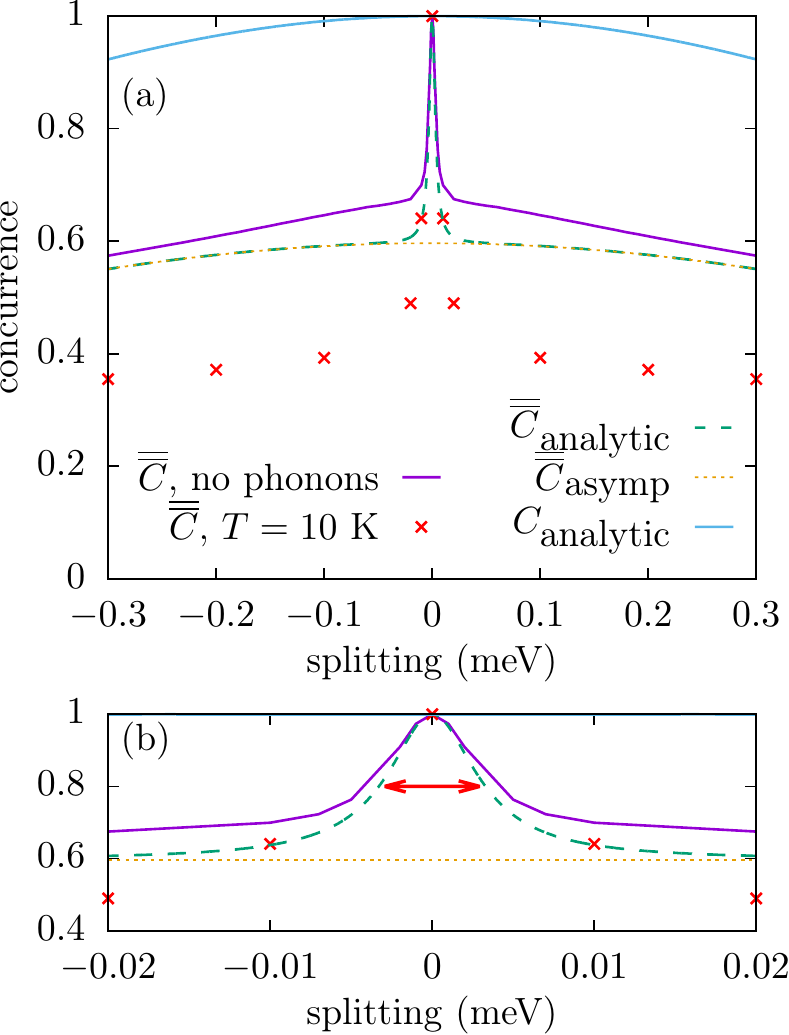}
\caption{(a) 
Double-time integrated concurrence 
$\overline{\overline{C}}$ as a function of the fine structure
splitting $\delta$ for the phonon-free case and  with phonons for a temperature $T=10$~K
compared with its analytic approximation $\overline{\overline{C}}_\textrm{analytic}$
in the phonon-free case
and the corresponding asymptotic behavior $\overline{\overline{C}}_\textrm{asymp}$.
Also shown is the analytic result  
$C_\textrm{analytic}$ for the single-time integrated concurrence without phonons.
(b) shows a zoom of the central peak where
the red double arrow  indicates a FWHM of 
$2\hbar\kappa \lambda^2={8\hbar^{3}\kappa g^{2}}/{E_{B}^{2}}$.
\label{fig:Cvss_nophon}}
\end{figure}

In Fig.~\ref{fig:Cvss_nophon}, the numerically calculated 
double-time integrated concurrence $\overline{\overline{C}}$ for the phonon-free case
is depicted as a function of the fine structure splitting $\delta$ for
$\hbar g=0.1$~meV, $\kappa=0.25$~ps$^{-1}$, and $E_B=1.5$~meV
and compared with the analytic expressions
for the time-dependent and double-time integrated concurrence.
The analytic expression for the double-time integrated concurrence reproduces
the main features of the numerical results quite well.
For further comparison we also show in Fig.~\ref{fig:Cvss_nophon} results for
$\overline{\overline{C}}$ accounting for phonons at a temperature $T=10$~K.
The necessary evaluation of two-time correlation functions in the presence of phonons
has been carried out by a numerically exact path integral approach, the details of which will
be discussed elsewhere.

Let us first concentrate on the results for the phonon-free case.
In contrast to our analytic expression $C_{\text{analytic}}$, which approximates well
the time-dependent concurrence $C(t)$ as seen before and
which remains close to one even for large fine structure splittings, the double-time
integrated concurrence $\overline{\overline{C}}$ typically has a 
narrow Lorentzian-like peak at splittings close to zero with a width 
$\sim 10$~\textmu eV
and, for larger splittings, reaches a plateau.
This qualitative behavior of the double-time integrated concurrence 
has been reported before in numerical studies \cite{Jahnke2012,Heinze2017}
and is also in agreement with experimental results \cite{Stevenson2006,Hudson2007,Hafenbrak}.
However, to the best of our knowledge, key quantities, such as the width of the 
central peak and the height of the plateau, have not yet been explained in 
terms of the microscopic parameters of the system.
Here, the derivation of the analytic expression for the
double-time integrated concurrence allows us to identify these key quantities.
In particular, recall that the numerator in the analytic expression for
the double-time integrated concurrence in Eq.~\eqref{eq:ooCanalyt} has
two contributions: the terms containing the parameter $p$ originating from
transitions through exciton-like energy 
eigenstates $\tilde{X}'_i$ of the four-level system in 
Eqs.~\eqref{eq:basis4lvl}, and another term from 
the transitions through the lowest energy eigenstates $\tilde{G}'_i$.
Because $p$ is proportional to the fine structure splitting $\delta$ 
and the contribution through the exciton-like eigenstates $\tilde{X}'_i$
decays for large values of $p$ as $\frac 1 p$, the double-time integrated
concurrence for large $\delta$ is determined by the contribution through 
the eigenstates $\tilde{G}'_i$. 
For $p\to\infty$, one obtains from the analytic expression in 
Eq.~\eqref{eq:ooCanalyt}
\begin{align}
&\overline{\overline{C}}_\textrm{asymp}=
C_\textrm{analytic} \frac{4g^2}{4g^2+\kappa^2}
=\frac{E_B^2-\delta^2}{E_B^2+\delta^2} \frac{4g^2}{4g^2+\kappa^2},
\label{eq:ooCasymp}
\end{align}
which is also plotted in Fig.~\ref{fig:Cvss_nophon}. 

As can be seen from the figure, the central peak can be attributed to the transitions
through the exciton-like eigenstates $\tilde{X}'_i$ and its width
is explained as follows: Due to the diagonalization of the four-level system,
the exciton-like states acquire a finite contribution from states involving
the ground state of the quantum dot and one cavity photon. The admixture
of these states leads to a mean photon number $n\approx\lambda^2_{H/V}$
for the states $\tilde{X}'_i$. Thereby the exciton-like eigenstates acquire
a loss term with a rate $\gamma\approx\kappa \lambda^2$. 
This loss rate $\gamma$ of an exciton-like eigenstate coincides 
with the cavity-assisted single-photon emission rate $\kappa_{1P}$ 
from the state $|X_H,0,0\rangle$ to $|G_H,0,0\rangle$ derived in 
Ref.~\onlinecite{delValle_twoPhoton}.
For a Lorentzian 
resonance, the full width at half maximum (FWHM) of the spectrum 
is related to the exponential decay rate $\gamma$ by 
FWHM=$2\hbar \gamma$, which in our case yields 
FWHM=$2\hbar\kappa \lambda^2={8\hbar^{3}\kappa g^{2}}/{E_{B}^{2}}$.
This value is indicated in the inset of Fig.~\ref{fig:Cvss_nophon} as 
a red double arrow and agrees well with the FWHM of the central peak
of the double-time integrated concurrence.
Finally, we note that at $\delta=0$ the double-time integrated concurrence
reaches its maximum value $\overline{\overline{C}}=1$ even when phonons are accounted
for and thus agrees in this case with $\overline{C}$. However, introducing a phenomenological
pure dephasing of the coherences between electronic configurations has been reported \cite{Jahnke2012,Heinze2017}
to result in lower values of $\overline{\overline{C}}$ at  $\delta=0$.

Phonons have a noticeable impact on the double-time integrated concurrence as can be seen, e.g.,
from the results (red crosses) shown in Fig.~\ref{fig:Cvss_nophon} for $T=10$~K.
In particular, phonons drastically reduce the concurrence for
larger splittings while for small $\delta\lesssim{} 0.01\,$~\textmu eV  there is
almost no phonon influence.  
Overall, qualitative trends, like the narrow peak of $\overline{\overline{C}}$
as a function of $\delta$ as well as the plateau obtained for larger splittings, 
remain similar to results obtained without accounting for phonons. 
This is in line with previous theoretical calculations
in Ref.~\onlinecite{Heinze2017} on the basis of master equations
in the polaron frame \cite{NazirReview}.

In the literature \cite{EdV,Pfanner2008} also the impact of frequency filtering of the
emitted photons on the behavior of concurrences evaluated for finite $\delta$
has been discussed. It is worthwhile to note that there are cross-relations 
between frequency filtering and selecting photons according to the
delay of their  emission. 
This is best understood by noting that the emission of the cavity 
tuned in resonance to the two-photon transition from the ground to the biexciton state
typically exhibits emission lines at the energies of the dipole-coupled dot transitions
as well as at the two-photon transition \cite{delValle_twoPhoton,EdV}.
Emissions via dipole-coupled dot transitions correspond to a cascaded decay where
first a single photon is emitted in a transition from the biexciton to one of 
the excitons and then, at a later time, a second photon is generated in the decay
of an exciton to the ground state. In contrast, the two photons
from the direct biexciton-to-ground state transition are generated
almost simultaneously with a much narrower spread in the emission time
than in cascaded emissions. 
Thus, filtering the emitted signal
at the frequency for the two-photon transition one collects a subset of
photons with a low spread in $\tau$ similar to measuring the single-time
integrated concurrence. Indeed, for a weakly driven cavity Ref.~\onlinecite{EdV}
reported in this case values near one independent on $\delta$ for the single- as well as the double-time
integrated concurrences. 
On the other hand, filtering at frequencies of the dipole-coupled transitions or between the
fine-structure split exciton lines results in low values for the double-time integrated
concurrence \cite{EdV,Pfanner2008} while the single-time integrated concurrence stays
close to one. Thus, using the already discussed argument of Stevenson \emph{et al.}, according to  
which simultaneously emitted photon pairs are expected to have higher degrees of entanglement
than photon pairs emitted with a delay, all tendencies observed for different frequency filtered
emissions can be nicely explained.

Also the role of the cavity can to some extent be discussed from the perspective that 
a cavity provides a frequency filter.
However, it should be noted that a cavity tuned to the ground-to-biexciton state transition
in general filters photons not only at the frequency of the two-photon transition but also 
at frequencies corresponding to photons emitted in a cascaded decay, as is evident from
the corresponding emission spectra \cite{delValle_twoPhoton,EdV}. The relative weights 
between two-photon and cascaded emissions is governed by the ratio of the respective
emission rates $\kappa_{2P}/\kappa_{1P}=4g^2/\kappa^2$. Thus, for weakly coupled cavities ($g\ll\kappa$)
the cavity essentially filters only the cascaded emission which should lead to low double-time
integrated concurrences. Indeed, we find from our analytic result Eq.~\eqref{eq:ooCanalyt}
\begin{align}
\lim_{g\to0}  \overline{\overline{C}}_\textrm{analytic}=
  \begin{cases}
    1, & \text{for } \delta=0\\
    0, & \text{otherwise}
  \end{cases},
\end{align}
i.e., a maximally sharp drop of the double-time integrated concurrence as a function of $\delta$.
Reaching the limit $g=0$, however, is for a cavity without driving a highly singular case since, for
vanishing $g$, there is no coupling between the dot levels and the system simply remains in the biexciton state
if radiative recombination is disregarded.
The above discussion therefore applies for small but finite $g$.
In the opposite limit $g\gg\kappa$, the assumption $g\ll E_{B}$ made in the derivation of our analytic result
may be violated so that, strictly speaking, Eq.~\eqref{eq:ooCanalyt} can no longer be used. 
Nevertheless, the tendency expected from our above discussion that,
in this limit, the cavity essentially filters only the simultaneous emission and thus the double-time integrated concurrence
should approach high values is corroborated by formally taking the limit $g\to\infty$ in 
Eq.~\eqref{eq:ooCanalyt}, which yields
\begin{align}
\lim_{g\to\infty}  \overline{\overline{C}}_\textrm{analytic} = \overline{C}_\textrm{analytic}
=\frac{E_B^2-\delta^2}{E_B^2+\delta^2}.
\end{align}
This is in accordance with our expectation as well as the results in Ref.~\onlinecite{EdV} for
an emission filtered at the two-photon transition.

\begin{figure}[t]
\includegraphics{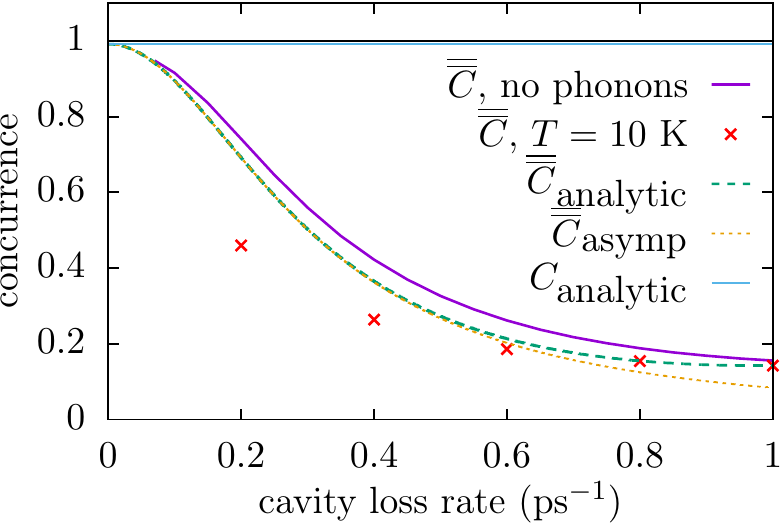} 
\caption{
Double-time integrated concurrence 
$\overline{\overline{C}}$ 
for a fine structure splitting $\delta=0.1$~meV
as a function of the cavity loss rate $\kappa$ without phonons
and with phonons at temperature $T$=10 K
compared 
with the analytic result  $\overline{\overline{C}}_\textrm{analytic}$ for the double-time integrated 
concurrence without phonons and its asymptotic behavior $\overline{\overline{C}}_\textrm{asymp}$.
Also shown is the analytic result $C_\textrm{analytic}$ for the time-dependent concurrence in the phonon-free case. 
\label{fig:Cvskappa}}
\end{figure}

Apart from the $\delta$ dependence also the impact of the cavity loss rate $\kappa$
on the different concurrences compared in this paper is instructive.
The dependence of the double-time integrated concurrence 
$\overline{\overline{C}}$ on  $\kappa$ is depicted with and without phonons
in Fig.~\ref{fig:Cvskappa} for the same fine structure splitting $\delta=0.1$~meV
as used in Fig.~\ref{fig:Cstivskappa} for the single-time integrated concurrence. 
While the analytic expression for the time-dependent concurrence 
$C_\textrm{anayltic}$ predicts a value close to one and 
independent of the loss rate, the double-time integrated concurrence
in Fig.~\ref{fig:Cvskappa}
significantly depends on $\kappa$ already without phonons: 
it decreases monotonically with increasing $\kappa$ and 
follows the asymptotic expression in Eq.~\eqref{eq:ooCasymp} for the 
height of the plateau. The latter is due to the fact that we are considering here 
a value of $\delta$ where in Fig.~\ref{fig:Cvss_nophon} already the plateau
is reached. While the unnormalized density matrix elements
entering the numerator and the denominator in the expression for the single-time integrated
concurrence are affected in the same way by a change of $\kappa$, this is not the
case for the double-time integrated concurrence. Here, the unnormalized density matrix
elements reflect, according to Eq.~\eqref{double-int-unnormalized}, the competition between
two-photon and sequential single-photon processes.  The cavity loss rate enters the corresponding 
contributions in two ways: First, the relative weight of two- and single-photon parts is
governed by the ratio of the corresponding emission rates
$\kappa_{2P}/\kappa_{1P}=4g^{2}/\kappa^{2} =\tilde\rho_{\tilde G_i\tilde G_j}(0)/\tilde\rho_{\tilde X_i\tilde X_j}(0)$.
Second, the two-photon parts decay as a function of $\tau$ without oscillations while
the sequential single-photon contributions exhibit oscillations reflecting the
relative phase between the two involved exciton components. This translates
after $\tau$ integration into prefactors $\sim 1/\kappa$
and $\sim1/[\frac 12\kappa (\lambda_i^2+\lambda_j^2)
-\frac i\hbar(E_{\tilde X'_i}-E_{\tilde X'_j})]$. Altogether, in the limit of vanishing $\kappa$
the two-photon contribution dominates irrespective of the other parameters due to the
$\sim1/\kappa^{3}$ singularity and $\overline{\overline{C}}_\textrm{analytic}$
approaches $C_\textrm{analytic}$ which is close to one. 
For small enough $\kappa$ and finite $\delta$ the
concurrence has reached the plateau with respect to its $\delta$ dependence
and is thus well described by Eq.~\eqref{eq:ooCasymp}, indicating a drop with
rising $\kappa$ following a Lorentzian with a FWHM $4g$. We note in passing that when
$\kappa$ is further increased for fixed other parameters the width of the peak in
Fig.~\ref{fig:Cvss_nophon} grows $\sim\kappa$ such that  
according to Eq.~\eqref{eq:ooCanalyt} $\overline{\overline{C}}_\textrm{analytic}$
recovers again to the high value $C_\textrm{analytic}$
in the limit $\kappa\to\infty$. However, for the parameters used in
Fig.~\ref{fig:Cvskappa} this recovery occurs for much larger $\kappa$ values than
covered in the plot.

For $\overline{C}$ as well as $\overline{\overline{C}}$ the interaction with
phonons leads to a reduction of the concurrence due to decoherence and because phonons cause the system to be in
a mixed state. As discussed before, the phonon impact decreases with increasing cavity loss rate because
higher $\kappa$ values limit the time-window over which the phonon-induced decoherence can take place.
For the double-time integrated concurrence this is nicely seen in Fig.~\ref{fig:Cvskappa} where  
the results with and without phonons approach each other for large $\kappa$.
Altogether we find a diminishing phonon influence for rising $\kappa$ on top of a nearly constant behavior of the single-time
integrated concurrence, while for the double-time integrated concurrence this is superimposed on a strong decay.
The resulting total effect is a complete trend reversal for finite $\delta$ in the presence of phonons: 
$\overline{C}$ increases with rising $\kappa$ while $\overline{\overline{C}}$ decreases
as long as $\delta$ is large enough to be in the plateau in Fig.~\ref{fig:Cvss_nophon}.

\begin{figure}[t]
\includegraphics{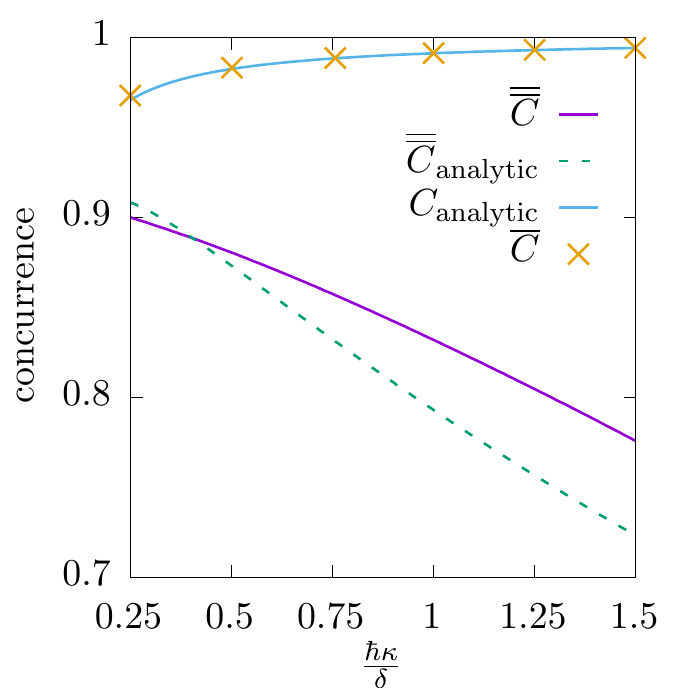}
\caption{Single- and double-time integrated concurrences for the phonon-free case plotted
vs.~$\frac{\hbar\kappa}{\delta}$ for fixed $\hbar\kappa\,\delta=0.01$~meV$^{2}$.
Shown are the numerically obtained results together with the corresponding analytic
approximations.
\label{fig:C1C2}}
\end{figure}

Interestingly, while the trend reversal as a function of the cavity loss rate
is attributed to the phonon influence,
the non-equivalence of $\overline{C}$ and $\overline{\overline{C}}$ can
already be demonstrated in the phonon-free case.
To this end, we plot $\overline{C}$ and $\overline{\overline{C}}$ 
as a function of $\frac{\hbar\kappa}{\delta}$ for fixed $\hbar\kappa\,\delta$ in Fig.~\ref{fig:C1C2}.
While $\overline{C}$ is increasing monotonically with rising
$\frac{\hbar\kappa}{\delta}$, $\overline{\overline{C}}$ is decreasing.
The observation of opposite trends in the single- and double-time integrated concurrences has important implications 
for the interpretation of the results. 
When comparing different situations, e.g., cavities with different values $\kappa_1$ and $\kappa_2$, it may turn out that 
according to $\overline{C}$ the cavity with $\kappa_1$ gives rise to the higher entanglement while for $\overline{\overline{C}}$ 
$\kappa_2$ leads to the higher entanglement or vice versa. 
This clearly demonstrates that $\overline{C}$ and $\overline{\overline{C}}$ cannot be equivalent measures for the same physical quantity, 
but instead measure two different types of entanglement.

\section{Conclusion}

We have studied the two-photon state generated by the biexciton cascade in a quantum dot
embedded in a microcavity.
Our main focus is a comparison of  dependencies of three different practical definitions of 
concurrences prevalent in the literature  on relevant parameters such as the exciton fine structure splitting
and the cavity loss rate.
In particular, we compare the time-dependent concurrence $C(t)$ associated with the state of the
system at a given time $t$, which by definition reflects the corresponding entanglement of formation, 
with concurrences derived from the results of different quantum state reconstruction strategies.
These strategies are based on photon coincidence measurements collecting data points
either over extended real- and delay-time intervals (double-time integrated concurrence $\overline{\overline{C}}$) or
over an extended real-time interval but a narrow delay-time window (single-time integrated concurrence
$\overline{C}$). Considering the photons recorded in the two detectors
of the coincidence measurements, these concurrences refer to photons with
a simultaneous emission time $t$ ($C(t)$), photons resulting from simultaneous emission irrespective of the time $t$ of
the emission ($\overline{C}$), and photons resulting from emissions without resolving the emission times of both photons 
($\overline{\overline{C}}$).

For a quantum dot with finite biexciton binding energy ($E_B \gg \delta$) in a cavity whose
modes are tuned in resonance with the two-photon transition between the ground
and the biexciton state, we have derived analytic expressions for the time-dependent, the single-time integrated,
and the double-time integrated concurrence in the absence of dot-phonon interaction.
Our results are applicable also beyond the weak-coupling limit of the dot-cavity system
and we have shown that they agree well with the results obtained from numerical calculations.

The single-time integrated concurrence, which can be accessed experimentally by 
time-windowing techniques\cite{StevensonPRL2008} or using vary narrow time bins for
the delay time\cite{Bounouar_2018}, is found to be close to the stationary value of 
the time-dependent concurrence at long times. This remains true
when the dot-phonon interaction is fully taken into account. 
The reason for this agreement between $\overline{C}$ and $C(t)$ is that,
due to the high energetic penalty involved in the occupation of 
single-exciton states with one cavity photon, 
the dynamics of the biexciton cascade is essentially incoherent and 
exponential, even when the coupling $g$ is comparable to the cavity loss rate $\kappa$. 
Because all oscillations between the states
with two photons in either the horizontally or the vertically polarized 
cavity mode are overdamped, the time-integration does not lead to 
a significant dephasing of coherences between those states.
Thus, in the dot-cavity configuration considered here, the information 
contained in $C(t)$ 
about the entanglement of formation assigned to two-photon states in the cavity at time $t$ 
is accessible by the more easily measurable single-time integrated
concurrence $\overline{C}$. 

In contrast, the double-time integrated concurrence, which in most experiments is the measured quantity
since measuring over extended real-time and delay-time intervals
provides the highest photon coincidence counts, shows a completely different 
behavior than either $C(t)$ or $\overline{C}$. 
First of all, disregarding phonons, $\overline{\overline{C}}$ as a function of the fine structure splitting
$\delta$ has a narrow peak with a FWHM of approximately $10$~\textmu eV and
then drops to a plateau that only weakly depends on the value of the splitting. 
We also obtain analytic expressions for the height of the plateau as well as 
the FWHM of the central peak and thereby explain the shape of 
the double-time integrated concurrence as a function of $\delta$.
Numerical simulations reveal that phonons do not change this behavior qualitatively
and only lead to quantitative corrections, which is in line with previous studies\cite{Heinze2017}.
In contrast, neither with nor without phonons the time-dependent concurrence
exhibits a narrow peak as a function of the fine structure splitting which 
evolves into a plateau for larger splittings.
In the phonon-free case it stays close to one even for splittings as large
as $\sim 0.1$~meV, while phonons lead to a reduction that follows
a bell-shaped curve.
This different behavior upon variation of the fine structure splitting can be attributed to the fact that, 
even in the presence of a non-vanishing fine structure splitting, the two emitted photons are strongly entangled 
at any given time, which is reflected in the time-dependent and the single-time integrated concurrence.
However, the integration over the delay time in the double-time integrated concurrence gives rise to destructive
interference of the two pathways.

Another qualitative difference between the time-dependent concurrence
and the double-time integrated concurrence is the dependence on the cavity loss rate: 
The analytic expression for $\overline{C}$, which does not account for the
effects of phonons, is independent of the cavity loss rate.
Taking dot--LA-phonon interactions into account using a
numerically exact real-time path-integral method \cite{PI_cQED} 
reveals that phonons can lead to a reduction of the concurrence
and that the concurrence increases with the cavity loss rate.
The reason for this is that the losses limit the time available for 
dephasing processes, so that for large $\kappa$ the phonon-induced
reduction of the concurrence is suppressed.
In contrast, for the considered situation $\overline{\overline{C}}$ decreases with
increasing loss rate in the phonon-free case as well as when phonons are taken into account.
This $\kappa$ dependence of $\overline{\overline{C}}$ essentially reflects 
the competition between two- and single-photon processes.

From these results it clearly follows 
that upon variation of the cavity loss rate 
opposite orderings are obtained with respect to
the two figures of merit  provided by
$\overline{C}$ and $\overline{\overline{C}}$, respectively.
In addition, already in the phonon-free case 
it turns out that $\overline{C}$ and $\overline{\overline{C}}$
exhibit opposite trends when varying the ratio between $\kappa$ and $\delta$
while keeping the product of these quantities constant.
Altogether, this implies that single- and double-time integrated 
concurrences cannot be equivalent measures for the same physical quantity
but instead reflect different aspects of entanglement.

\section*{Acknowledgments}
\label{sec:Acknowledgments}

M. Cygorek thanks the Alexander-von-Humboldt foundation for support through a Feodor Lynen fellowship
while A.M. Barth and V.M. Axt gratefully acknowledge financial support from Deutsche Forschungsgemeinschaft via the
Project No. AX 17/7-1.

\appendix
\section{Dynamics in the effective three-level system\label{app:3lvl}}

Here, we calculate the dynamics in the weakly coupled three-level system 
spanned by the states $|G'_H\rangle,|G'_V\rangle$ and $|B'\rangle$ defined in 
Eqs.~\eqref{eq:basis3lvl}.
To this end, we define the density matrix elements 
\begin{align}
&\rho_{\nu'\mu'}:=\langle \big(|\nu'\rangle\langle \mu'|\big)\rangle,
\end{align}
with $\nu',\mu'\in\{G'_H,G'_V,B'\}$.

In the subspace spanned by the above states, the Lindblad losses
induce transitions to states with lower excitation numbers.
However, these states can be disregarded in the equation of motion
for $\rho_{\nu'\mu'}$ since they do not couple back to states with 
higher excitation numbers.
Thus, the trace of the density matrix $\rho$ in the subspace spanned by
$\{G'_H,G'_V,B'\}$ is no longer conserved and we find
\begin{align}
\ddt \rho_{\nu'\mu'}=&\frac i\hbar \sum_{\bar\nu} 
[H'_{\nu'\bar\nu} \rho_{\bar\nu\mu'}- \rho_{\nu'\bar\nu}H'_{\bar\nu\mu'}] \nn
&
-\frac \kappa2\sum_{\bar\nu}( n_{\nu'\bar\nu} \rho_{\bar\nu\mu'}
+ \rho_{\nu'\bar\nu}n_{\bar\nu\mu'}),
\end{align}
where $n_{\nu'\mu'}=\sum_{i=H,V} (a^\dagger_i a_i)_{\nu'\mu'}$
is the photon number operator. 
The equations of motion for the effective three-level system are then
\begin{subequations}
\begin{align}
\label{eq:eom3lvlBB}
\ddt \rho_{B'B'}=&-(\lambda_H^2+\lambda_V^2)\kappa \rho_{B'B'}
- \!\!\! \sum_{i\in\{H,V\}} \!\!\! 2\tilde{g}_i \Im(\rho_{G'_i B'}), \\
\label{eq:eom3lvlGB}
\ddt \rho_{G'_i B'}=&\Big(i \frac{\tilde\delta_i}\hbar-\kappa\Big)\rho_{G'_iB'}
+i\tilde{g}_i\rho_{B'B'} - \!\!\! \sum_{j\in\{H,V\}} \!\!\! i\tilde{g}_{j}\rho_{G'_iG'_j},\\
\label{eq:eom3lvlGG}
\ddt \rho_{G'_iG'_j}=& \Big[ i 
\Big(\frac{\tilde{\delta}_i}\hbar - \frac{\tilde\delta_j}\hbar\Big)
-2\Big(1-\frac 12(\lambda_i^2+\lambda_j^2)\Big) \kappa\Big]\rho_{G'_iG'_j}\nn&
+i\tilde{g}_i \rho_{B'G'_j}-i\tilde{g}_j\rho_{G'_i B'},
\end{align}
\label{eq:eom3lvl}
\end{subequations}
where we have defined $\tilde{\delta}_{H/V}=\pm \tilde\delta$.
It is straightforward to see that when the system is initially 
in the biexciton state, $\rho_{G'_iB'}=\mathcal{O}(\lambda)$ and
$\rho_{G'_iG'_j}=\mathcal{O}(\lambda^2)$.
Furthermore, we consider the weak-coupling regime in the effective 
three-level system where $\tilde{g}_i\ll \kappa$. Therefore, 
$\rho_{G'_iG'_j}$ decays fast compared to $\rho_{B'B'}$ due to the losses
and can be neglected for the calculation of $\rho_{G'_iB'}$.
Then, the coherences $\rho_{G'_i B'}$ are given by
\begin{align}
\rho_{G'_i B'}(t)=i\tilde{g}_i \int\limits_0^t dt'\,
e^{(-\kappa+i\tilde\delta_i/\hbar)(t-t')} \rho_{B'B'}(t').
\end{align}
Because $\rho_{B'B'}$ changes only on a much longer time scale 
(all terms on the r.h.s. of Eq.~\eqref{eq:eom3lvlBB} are of the order
$\mathcal{O}(\lambda^2)$) than $\rho_{G'_iB'}$, one can apply the Markov
limit consisting of evaluating $\rho_{B'B'}(t')$ at $t'=t$ and
setting the lower limit of the intergral to $-\infty$, so that
\begin{align}
\rho_{G'_i B'}(t)\approx i\frac{\tilde{g}_i}
{\kappa-\frac i\hbar\tilde{\delta}_i} \rho_{B'B'}(t).
\label{eq:rhoGBfromBB}
\end{align}
Feeding this result back into the equation for $\rho_{B'B'}$ and dropping
terms higher than second order in $\lambda$, one finds
\begin{align}
&\rho_{B'B'}(t)\approx e^{-\kappa_B t}, \\
&\kappa_B\approx(\lambda_H^2+\lambda_V^2)\Big(\frac{4g^2}{\kappa}
+\kappa \Big).
\end{align}
Using again Eq.~\eqref{eq:rhoGBfromBB} one obtains explicit expressions 
for $\rho_{G'_i B'}$ and its complex conjugate 
$\rho_{B' G'_i}=(\rho_{G'_i B'})^*$, which are the source terms necessary
for the calculation of $\rho_{G'_iG'_j}$ from Eq.\eqref{eq:eom3lvlGG}:
\begin{align}
\rho_{G'_iG'_j}(t)=&
\left(\frac{\tilde g_i\tilde g_j}{\kappa+\frac i\hbar\tilde\delta_j}
+\frac{\tilde g_i\tilde g_j}{\kappa-\frac i\hbar\tilde\delta_i}\right) \nn&
\times \int\limits_0^t dt'\,
e^{[-(2-\lambda_i^2-\lambda_j^2)\kappa +i(\tilde\delta_i-\tilde\delta_j)/\hbar]
(t-t')}  e^{-\kappa_B t'}.
\end{align}
Integrating over $t'$ and keeping only terms up to second order in the 
prefactor yields
\begin{align}
&\rho_{G'_iG'_j}=\frac{\tilde{g}_i\tilde{g}_j}{\kappa^2}
\Big(e^{-\kappa_B t}-e^{[-(2-\lambda_i^2-\lambda_j^2)\kappa
+i(\tilde\delta_i-\tilde\delta_j)/\hbar]t} \Big).
\label{eq:rhoGG3lvl}
\end{align}

\bibliography{PIbib}

\end{document}